\documentclass[10pt,
showpacs,
preprintnumbers,amsmath,amssymb,
aps,prd,nofootinbib,eqsecnum,a4paper]{revtex4}
\usepackage{epsfig}
\usepackage{graphicx,epsf}
\usepackage{color}
\usepackage{bm}
\usepackage{psfrag}
\usepackage[symbol]{footmisc}
\usepackage{tensor}
\def\be{\begin{equation}}
\def\ee{\end{equation}}
\def\bea{\begin{eqnarray}}
\def\eea{\end{eqnarray}}
\def\bi{\begin{itemize}}
\def\ei{\end{itemize}}
\def\p{\partial}

\def\H{{\cal H}}

\def\P{{\cal P}}
\def\cs2{c_{\rm{s}}^2}
\newcommand\eq[1]{Eq.~(\ref{#1})}
\newcommand\eqs[1]{Eqs.~(\ref{#1})}
\def\dd{\text{d}}
\def\drf{\delta^{(1)} \rho}
\def\drs{\delta^{(2)} \rho}
\def\dkf{\delta^{(1)} k}

\def\dRf{\delta^{(1)} R}
\def\dRs{\delta^{(2)} R}
\def\dzf{\delta^{(1)} z}
\def\dzs{\delta^{(2)} z}
\def\dVf{\delta^{(1)} V}
\def\dVs{\delta^{(2)} V}
\def\daf{\delta^{(1)} d_{A}}
\def\das{\delta^{(2)} d_{A}}
\def\E{{\cal E}}
\def\Ef{\delta^{(1)}{\cal E}}
\def\Es{\delta^{(2)}{\cal E}}
\def\dS{\delta^{(1)}\Sigma}
\def\dnuf{{\delta^{(1)} \nu}}
\def\dnf{{\delta^{(1)} n}}
\def\dnus{{\delta^{(2)} \nu}}
\def\dns{{\delta^{(2)} n}}
\def\ff{{\Phi_{1}}}
\def\pf{{\Psi_{1}}}
\def\fs{{\Phi_{2}}}
\def\ps{{\Psi_{2}}}
\def \beg {\begin{enumerate}}
\def \en {\end{enumerate}}

\def\M0{{\cal M}_0}

\def\H{\mathcal{H}}

\begin{document}

\title{Galaxy number counts at second order: an independent approach}

\author{Jorge L.~Fuentes$^{1}$ \footnote[3]{\href{mailto:j.fuentesvenegas@qmul.ac.uk}{j.fuentesvenegas@qmul.ac.uk}} }
\author{Juan Carlos~Hidalgo$^{2}$ \footnote[2]{\href{mailto:hidalgo@icf.unam.mx}{hidalgo@icf.unam.mx}}}
\author{Karim A.~Malik$^{1}$ \footnote[1]{\href{mailto:k.malik@qmul.ac.uk}{k.malik@qmul.ac.uk}}}

%
\affiliation{
$^1$Astronomy Unit, School of Physics and Astronomy, Queen Mary University of London, Mile End Road, London, E1 4NS, United Kingdom\\
$^2$Instituto de Ciencias F\'{i}sicas, Universidad Nacional Aut\'{o}noma de M\'{e}xico,\\Av. Universidad S/N. Cuernavaca, Morelos, 62251, M\'{e}xico}
\date{\today}

\begin{abstract}
Next generation surveys will be capable of determining cosmological parameters beyond percent level. To match this precision, theoretical descriptions should look beyond the linear perturbations to approximate the observables in large scale structure. A quantity of interest is the Number density of galaxies detected by our instruments. This has been focus of interest recently, and several efforts have been made to explain relativistic effects theoretically, thereby testing the full theory. However, the results at nonlinear level from previous works are in disagreement. We present a new and independent approach to computing the relativistic galaxy number counts to second order in cosmological perturbation theory. We derive analytical expressions for the full second order relativistic observed redshift, for the angular diameter distance and for the volume spanned by a survey. Finally, we compare our results with previous works which compute the general distance-redshift relation, finding that our result is in agreement at linear order.
\end{abstract}

\pacs{98.80.Cq \hfill  arXiv:1908.08400}

\maketitle

\section{Introduction}

Recent years have witnessed the beginning of the era of precision cosmology with future surveys such as BOSS \cite{BOSS}, eBOSS \cite{eBOSS}, Euclid \cite{euclid}, MeerKAT,  SKA, LSST \cite{santos2017meerklass, maartens2015cosmology, lsst2009},and WFIRST \cite{WFIRST} improving and tightening constraints on observable cosmological parameters. Additionally, great theoretical advancements have been made in tackling nonlinear regimes to test cosmological models and general relativity.

For theoretical cosmological models, probing the relation between redshift and angular diameter or luminosity distance of a source is of significant value.
This relation determines the parameters of the cosmological model, but when perturbations due to structure are included, new effects are revealed. One of these effects is lensing, which is observed along the line of sight. Another effect is the distortion in redshift space due to velocities and motion of the sources, giving rise to `Doppler lensing'. The integrated Sachs-Wolfe (ISW) effect which arises from integrating along the full line of sight between the source and the observer.

Most of the known effects on the distance-redshift relation are calculated at linear order in cosmological perturbation theory in Refs.~\cite{jeong1, durrer1, antony, chen, zalda}. However, at second order other general relativistic effects must be considered. When structure is evolving, nonlinear modes come into play, and many of these go beyond Newtonian theory.

One of the main observables directly affected by the angular and luminosity distance estimation is the galaxy number density (dubbed often as number counts). Important examples of these effects have been calculated in Refs.~\cite{cc1,cc2, durrer1, durrer2, yoo1, yoo2, yoo3, marozzi1, marozzi2, marozzi3}. The dominating terms of the full second order calculations have been reviewed in Ref.~\cite{nielsen}. More recently in Ref.~\cite{yoo5} the authors present second order relativistic corrections to the observable redshift. And even a ``pedagogical'' approach to the lengthy calculations is provided in Ref.~\cite{yoo4} to try to ease the tension between the different groups.

In this work, we present a new path to compute the second-order galaxy number in a Friedmann-Lema\^{i}tre-Robertson-Walker (FLRW) universe. This follows the volume determination as defined in Ref.~\cite{ellis} instead of computing the luminosity distance as in Ref.~\cite{yoo2}. We identify key effects, some of which will be observable with the next generation of cosmological surveys. To check the robustness of our results we confirm the consistency for the first order expressions with previous works.

This paper is organised as follows: In section \ref{sec:whatwemeasure} we provide all the definitions needed for the linear and nonlinear calculations in the context of cosmological perturbation theory (CPT). In section \ref{sec:effects} we compute the linear and nonlinear parts of the null geodesic equation, the observed redshift, and show the geometrical effects present at this level. In section \ref{sec:distances} we compute the angular diameter distance and the physical volume that the galaxy survey spans, only in this section we make a conformal transformation that maps null geodesics from the perturbed FLRW metric to a perturbed Minkowski spacetime $\hat{g}_{\mu \nu}\rightarrow g_{\mu \nu}$, how quantities transform under this map is discussed in further detail within this section. In section \ref{sec:counts} we compute our main result, the galaxy number overdensity. In section \ref{sec:comparison} we make a check for the calculation performed in this paper with other results in the literature at linear order and find an exact agreement with all of them pertaining the right interpretation of variables. Finally, in section \ref{sec:discussion} we give a discussion of our result, some conclusions and future work.

{\it{Notation.}} We use indices $\mu,\nu,\dots = 0,1,2,3$ in a general spacetime. In perturbed FLRW, the indices $i,j,\dots = 1,2,3$ denote spatial components. The derivative with respect to the conformal time is given by a dash
\be
\frac{\p X}{\p\eta} = X'.
\ee 
We use the notation
\be
\label{eq:not1}
\left(X\right)^{s}_{o} = X\big|^{s}_{o} = X_{s}-X_{o} = X(\lambda_{s})-X(\lambda_{o}).
\ee
The derivative with respect to the affine parameter is  
\begin{equation}
\label{eq:der-affine}
\frac{\dd X}{\dd \lambda} = X' + n^{i}X_{,i},
\end{equation}
where $n^{i}$ represents the direction of observation. This last equation implies, for a scalar function $X$,
\begin{equation}
\label{eq:double-der-affine}
n^{i}n^{j}X_{,ij}= n^{i}n^{j}\p_{i}\p_{j}X = \frac{\dd^{2} X}{\dd \lambda^{2}}-2 \frac{\dd X'}{\dd \lambda} +X'',
\end{equation}
where $\nabla_{i}X = \partial_{i}X = X_{,i}$ is the spatial part of the covariant derivative. 

\section{Basis for the definition of the galaxy number density}
\label{sec:whatwemeasure}

\subsection{Metric perturbations}
The perturbed FLRW spacetime is described in the longitudinal gauge by \cite{malik}
\be
\label{eq:metric}
\dd s^{2} = a^{2}\left[ -\left(1+2\ff+\fs \right) \dd \eta^{2} + \left( 1 - 2 \pf - \ps \right) \delta_{ij} \dd x^{i}\dd x^{j}\right],
\ee
where $\eta$ is the conformal time, $a=a(\eta)$ is the scale factor and $\delta_{ij}$ is the flat spatial metric, and we have neglected the vector and tensor modes, we also allow for first and second order anisotropic stresses. From now on we consider perturbations around a FLRW metric up to second-order.

\subsection{Matter velocity field and peculiar velocities}
The components of the 4-velocity $u^{\mu}=\dd x^{\mu}/\dd \eta$ up to and including second order using the perturbed metric are given by
\begin{align}
\label{eq:pertu0}
u_{0} &= -a \left[ 1 + \ff + \frac{1}{2} \fs- \frac{1}{2}\ff^{2}+\frac{1}{2}v_{1 k}v_{1}^{k} \right], \\
\label{eq:pertui}
u_{i} &= a \left[ v_{1 i} + \frac{1}{2} v_{2 i}  - 2 \pf v_{1 k}\right], \\
u^{0} &= a^{-1} \left[1-\ff - \frac{1}{2}\fs+\frac{3}{2} \ff^{2}+\frac{1}{2}v_{1 k}v_{1}^{k} \right],\\
u^{i} &= a^{-1} \left[ v_{1}^{i} + \frac{1}{2} v_{2}^{i} \right],
\end{align}
where $v_{i} = \partial_{i} \text{v}$, with $\text{v}$ the velocity potential.

\subsection{Photon wavevector}
\begin{figure}
\centering
\includegraphics[scale=1]{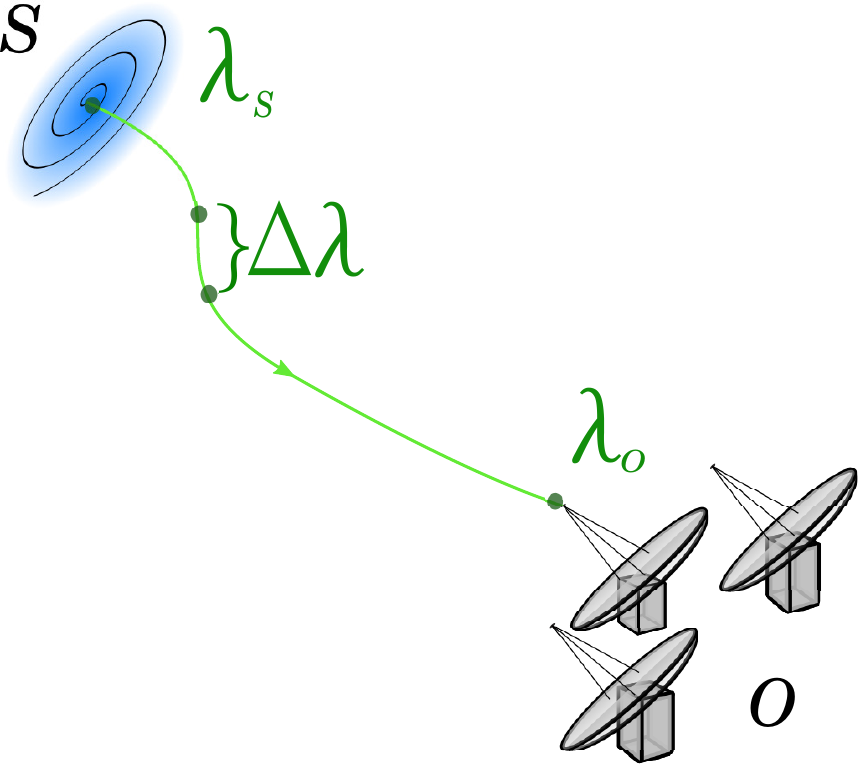}
\caption{Affine parameter convention of a light ray in a radio observation. $S$ denotes the source, $O$ is the observer and $\lambda$ is the affine parameter.}
\label{fig:affine}
\end{figure}

In a redshift survey galaxy positions are identified by measuring photons produced at the source, denoted by $s$,  and detected by an observer labelled $o$. In a general spacetime, we consider a lightray with tangent vector $k^{\mu}$ and affine parameter $\lambda$, that parametrises the curve the lightray follows. The source, $\lambda_{s}$, and the observer, $\lambda_{o}$, are represented by given values for the affine parameter, as illustrated in Fig.~\ref{fig:affine}. The components of the photon wavevector can be written as
\be
\label{eq:kbg}
\bar{k}^{\mu} = \frac{\dd x^{\mu}}{\dd \lambda} = a^{-1}\Big[1,n^{i}\Big],
\ee
where the overbar denotes background quantities, $n^{i}$ is the direction of observation\footnote{Some authors define $n^{i}$ with the opposite sign. See, for example, Refs.~\cite{durrer1,cc1,cc2}.} pointing from the observer to the source, and following the normalisation condition: $n^{i}n_{i} = 1$.

The tangent vector is \textit{null} 
\be
\label{eq:null}
k_{\mu}k^{\mu} = 0,
\ee
and \textit{geodesic} 
\be
\label{eq:geodesic}
k^{\nu}\nabla_{\nu}k^{\mu} =0.
\ee
where $\nabla_{\nu}$ is the covariant derivative defined by the metric given in \eq{eq:metric}.
In general, the perturbed wavevector can be written as
\be
\delta^{(n)} k^{\mu}=a^{-1}\Big[\delta^{(n)} \nu, \delta^{(n)} n^{i}\Big]. 
\ee
where $\delta^{(n)}$ gives the $n\text{-th}$ order perturbation, and we are following the usual notation for the temporal component, that is $k^{0}\equiv\nu$ \cite{yoo4}.

The affine parameter of the geodesic equation is also related to the comoving distance ($\chi$) by
\be
\label{eq:comovingd}
\chi = \lambda_{o}-\lambda_{s},
\ee
and in terms of the redshift this is
\be
\label{eq:comd}
\chi(z) = \int_{0}^{z} \frac{\dd \tilde{z}}{(1+\tilde{z})\H (\tilde{z})}.
\ee

\subsection{Observed redshift}

The photon energy measured by an observer with 4-velocity $u^{\mu}$ is 
\be
\label{eq:photon-energy}
\E = - g_{\mu \nu}u^{\mu}k^{\nu}.
\ee
From \eq{eq:photon-energy} the observed redshift of a source (e.g.~a galaxy) can be defined as
\be
\label{eq:redshift}
1+z = \frac{\E_{s}}{\E_{o}} ,
\ee

\noindent From this definition there will be a Doppler effect on the redshift due to the velocities $u^{\mu}$ and the observed redshift is in fact a function of the velocity and the wavevector, i.e.~$z=z(k^{\mu},u^{\mu})$.

\subsection{Angular diameter distance}

\begin{figure}
\centering
\includegraphics[trim=40 440 70 75,clip,scale=0.9]{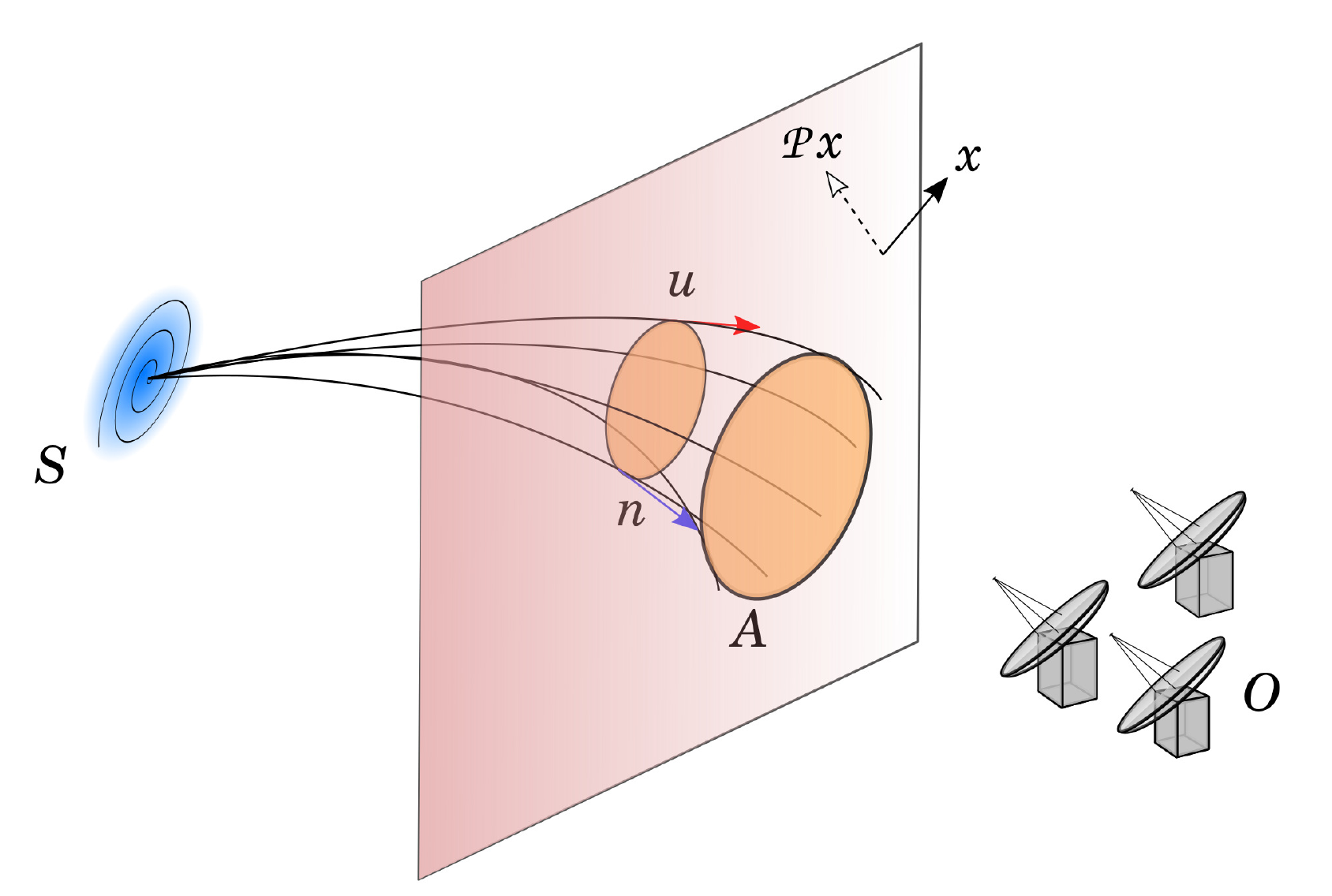}
\caption{The lightray bundle going from the source, $S$, to the observer, $O$, the cross sectional area created by the infinitesimal separation of the lightrays and the screen space, $A$, which is orthogonal to the 4-velocity, $u^{\mu}$, and the direction of observation, $n^{\mu}$, in this figure, for simplicity, it is shown $n^{i}$ pointing from the source to the observer, i.e.~as the opposite from the one using in all our calculations. We present a general 4-vector $x^{\nu}$ and its projection onto screen space $\P_{\mu \nu}x^{\nu}$.}
\label{fig:screen}
\end{figure}

\begin{figure}
\centering
\includegraphics[trim=80 420 120 140,clip,scale=0.9]{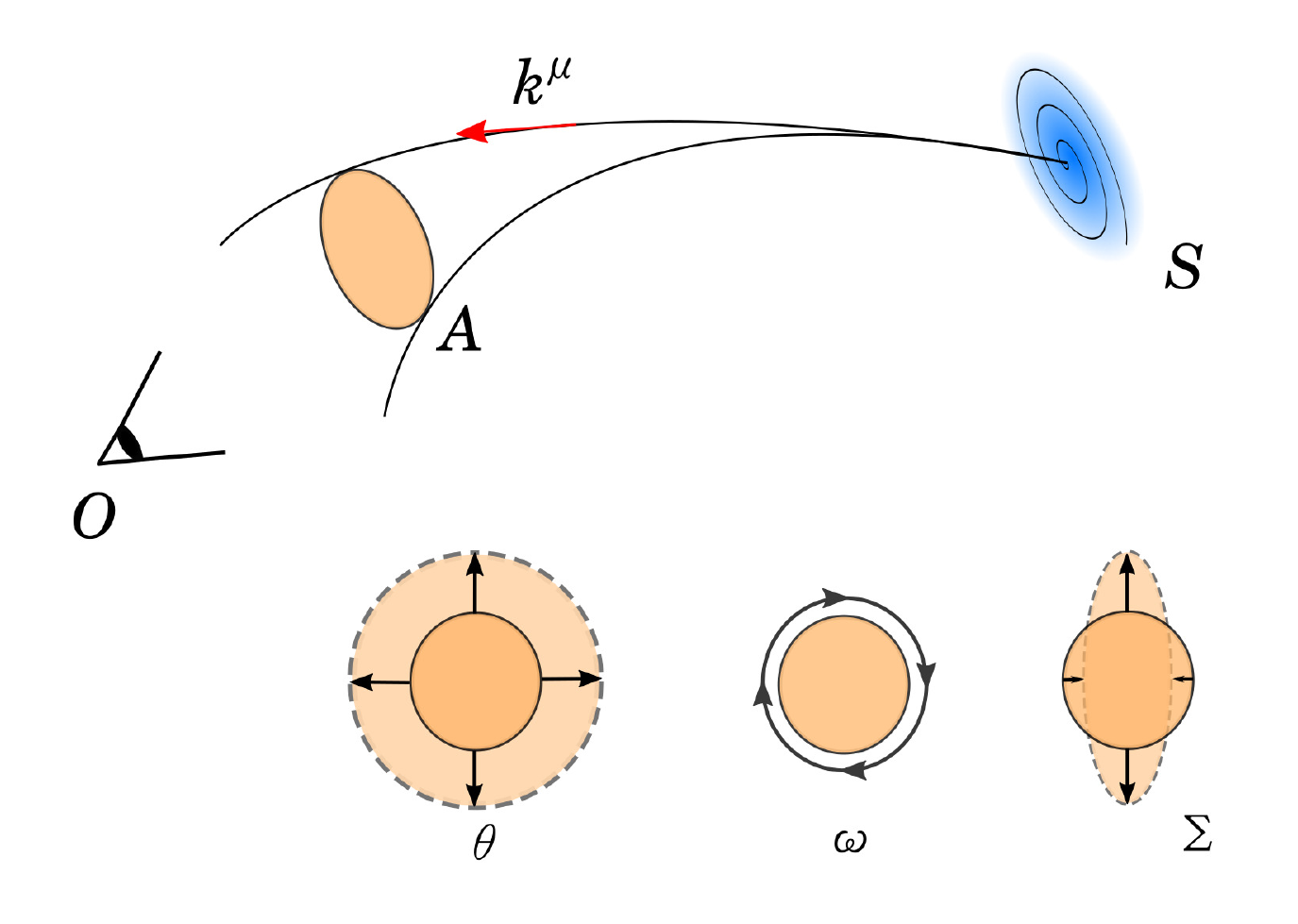}
\caption{ When a light-ray bundle travelling from a source, $S$, to an observer, $O$, passes close to matter, the cross section that the different null geodesics generate, $A$, gets distorted in different ways, and those are explained by an an expansion $\theta$, a vorticity $\omega$, and shear $\Sigma$, defined in Eqs.~\eqref{eq:deftheta} and \eqref{eq:defsigma}. The circles represent the area of the bundle's cross section and the arrows show how it is distorted.}
\label{fig:da}
\end{figure}

For a given bundle of lightrays leaving a source, the bundle will invariantly expand and create an area in between the lightrays that conform it, this area can be projected to a screen space, perpendicular to the trajectories of the photons and the 4-velocity of the observer, as illustrated in Figs.~\ref{fig:screen} and \ref{fig:da}.

The area of a bundle in screen space, $\mathcal{A}$, defines the angular diameter distance $d_{A}$, and is directly related to the null expansion $\theta$ \cite{cc4} defined in section \ref{sec:distances},
\begin{equation}
\label{eq:daa}
\frac{1}{\sqrt{\mathcal{A}}}\frac{\dd \sqrt{\mathcal{A}}}{\dd \lambda} = \frac{\dd \ln d_{A}}{\dd \lambda} = \frac{1}{2} \theta.
\end{equation}
where $\lambda$ is the affine parameter defined in \eq{eq:kbg}. Using \eq{eq:daa} we can compute how the area of the bundle changes along the geodesic trajectory that the photons are following from the source towards the observer.

\subsection{Physical Volume}

Number counts relate to the number of sources detected in a bundle of rays, for a small affine parameter displacement $\lambda$ to $\lambda+\dd\lambda$ at an event $P$. This corresponds to a physical distance 
\be
\label{eq:dl}
\dd \ell = (k^{\mu}u_{\mu}) \dd \lambda,
\ee 
in the rest frame of a comoving galaxy at said point in space $P$, if $k^{\mu}$ is a tangent vector to the past directed null geodesics (so that $k^{\mu}u_{\mu}>0$). 

\begin{figure}
\centering
\includegraphics[scale=0.7]{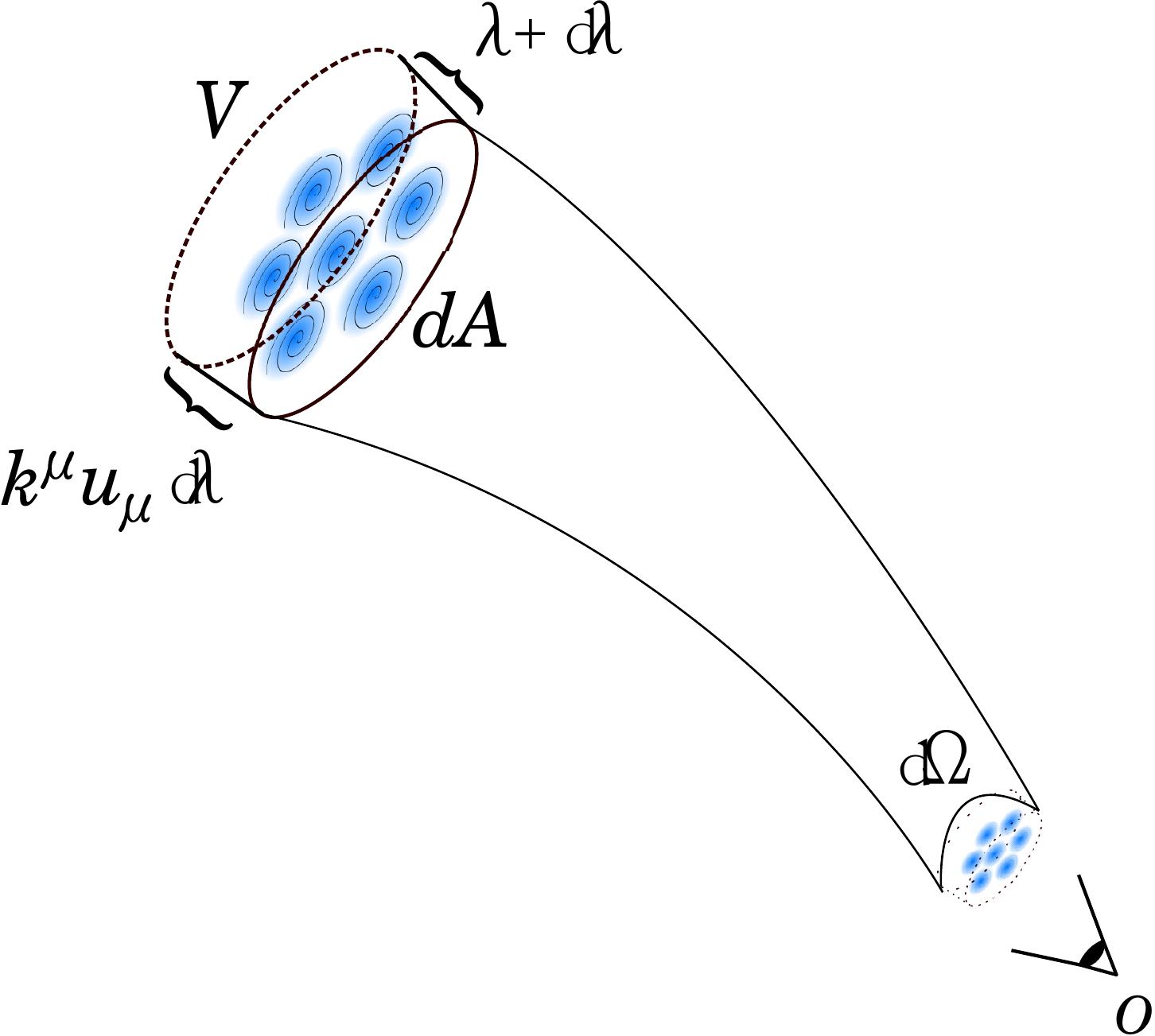}
\caption{Volume corresponding to an infinitesimal change in the affine parameter from $\lambda$ to $\lambda+\dd\lambda$.}
\label{fig:volume}
\end{figure}

The cross-sectional area of the bundle is 
\be
\label{eq:cross-a}
\dd \mathcal{A} = d_{A}^{2}(\lambda)\dd \Omega,
\ee 
if the geodesics subtend a solid angle $\dd \Omega$ at the observer, this is shown in Fig.~\ref{fig:volume}.

From \eqs{eq:dl} and \eqref{eq:cross-a} the corresponding volume element at a point $P$ in space is (see e.g.~\cite{ellis})
\begin{equation}
\label{eq:dldo}
\dd V = \dd \ell \dd \mathcal{A} = (k^{\mu}u_{\mu})d_{A}^{2}(\lambda) \dd \lambda \dd \Omega = -\E d^{2}_{A}(\lambda) \dd \lambda \dd \Omega.
\end{equation}

These covariant definitions lead to the expressions we compute in the following sections at first and second order in cosmological perturbation theory. 

\section{Perturbed null geodesics and redshift}
\label{sec:effects}

\subsection{Geodesic equation}

Let us now look at solutions to the geodesic equation. First, from \eq{eq:null} and the normalisation of $n^{i}$ we obtain the null condition of the photon wavevector
\begin{align}
\label{eq:null-conditions}
0 =& 2\Big[ n^{i}\dnf_{i} - \dnuf - \ff - \pf \Big] \notag \\
\qquad &+ \Bigg[ n^{i} \dns_{i} - \dnus - \frac{1}{2}\fs - \frac{1}{2} \ps - \left(\dnuf\right)^{2} + \dnf_{i} \dnf^{i} \notag\\
&\qquad\qquad\qquad\qquad\qquad\qquad - 4 \left(\ff+\pf\right) \dnuf - 4 \pf \left(\ff+\pf\right)\Bigg].
\end{align}
For the geodesic equation, \eq{eq:geodesic}, we get the propagation equations for the temporal and spatial perturbations 
\begin{align}
\label{eq:geo0}
k^{\mu}\nabla_{\mu}\delta^{(n)}\nu &= \frac{\dd \delta^{(n)}\nu}{\dd \lambda} + \Gamma^{0}_{\alpha \beta}k^{\alpha}k^{\beta}=0, \\
\label{eq:geoi}
k^{\mu}\nabla_{\mu}\delta^{(n)}n^{i} &= \frac{\dd \delta^{(n)}n^{i}}{\dd \lambda} + \Gamma^{i}_{\alpha \beta}k^{\alpha}k^{\beta}=0,
\end{align}
where $\Gamma^{\mu}_{\nu\sigma}$ are the connection coefficients at any given order $(n)$. Their expansion up to second order is provided in Appendix \ref{connection}.
Substituting and rearranging, we find the geodesic equations in general at first order \cite{durrer3},
\begin{align}
\label{eq:dnu1}
\frac{\dd \dnuf}{\dd \lambda} &= - 2 \frac{\dd \ff}{\dd \lambda} + \ff' + \pf',\\
\label{eq:dn1}
\frac{\dd \dnf^{i}}{\dd \lambda} &= 2 \frac{\dd \pf }{\dd \lambda}n^{i} -\left[\pf_{,}^{i} + \ff_{,}^{i}\right],
\end{align}
where \eq{eq:dnu1} there is a term related to the Integrated Sachs-Wolfe effect defined below. The expressions for the solution at second order are given in Appendix \ref{metriccomp}.
\subsection{Observed redshift}

We now expand the photon energy $\E = - g_{\mu \nu}u^{\mu}k^{\nu}$ up to second order,
\be
\E = \bar{\E} + \Ef + \frac{1}{2} \Es.
\ee
Using \eqs{eq:metric}, \eq{eq:pertu0}, \eqref{eq:pertui}, \eqref{eq:dnu1}, \eqref{eq:dn1}, \eqref{eq:dnu2}  and \eqref{eq:dni2}, we find that $\bar{\E}=1$, and
\begin{align}
\label{eq:dE1}
\Ef &= \dnuf + \ff - \left(v_{1 i}n^{i}\right), \\
\label{eq:dE2}
\Es &= \dnus + \fs - \left( v_{2 i}n^{i}\right) + 2 \ff \dnuf - 2v_{1i} \dnf^{i} + \ff^{2} + \left(v_{1 k}v_{1}^{k}\right) + 4 \pf \left( v_{1i}n^{i}\right) .
\end{align}

The perturbations to the photon energy given in \eqs{eq:dE1} and \eqref{eq:dE2} are written explicitly in terms of the metric potentials in Appendix \ref{metriccomp}.
Integrating \eqs{eq:dnu1}, \eqref{eq:dn1}, \eqref{eq:dnu2} and \eqref{eq:dni2}, with respect to the affine parameter, $\lambda$, from the observer to the source, we find the perturbed photon vector, $\delta k^{\mu}$. Note that $\delta k^{\mu}|_{o}=0$, what can be seen explicitly in \eq{eq:dE1-metric} and \eq{eq:dE2-metric},  so that
\begin{align}
\label{eq:dE1o}
\Ef\big|_{o} &= \ff |_{o} - (v_{1 i}n^{i})_{o}, \\
\label{eq:dE2o}
\Es\big|_{o} &= \fs |_{o} - \left( v_{2 i}n^{i}\right)_{o} + \left(\ff|_{o}\right)^{2} + \left(v_{1 k}v_{1}^{k}\right)_{o} + 4 \pf|_{o} \left( v_{1i}n^{i}\right)_{o}. 
\end{align}
The observed redshift is given by \eq{eq:redshift}, then up to second order that is
\be
\label{eq:redshift1}
1+z = 1 +\bar{z}+ \dzf+ \frac{1}{2} \dzs= \frac{\E_{s}}{\E_{o}} = \frac{(\bar{\E} + \Ef + \frac{1}{2} \Es)_{s}}{(\bar{\E} + \Ef + \frac{1}{2} \Es)_{o}}.
\ee
Note that in general, the expansion of the perturbed quotient up to second order is
\be
\label{eq:quotient}
\frac{A+\delta^{(1)} A + \frac{1}{2} \delta^{(2)}A}{B+\delta^{(1)} B + \frac{1}{2}\delta^{(2)}B} = \frac{A}{B}\left( 1 + \frac{\delta^{(1)} A}{A} - \frac{\delta^{(1)}B}{B} + \frac{\delta^{(2)}A}{2A} -\frac{\delta^{(2)}B}{2B} - \frac{\delta^{(1)}A}{A} \frac{\delta^{(1)} B}{B} + \left[\frac{\delta^{(1)}B}{B}\right]^{2}\right),
\ee
Using \eqs{eq:redshift1} and \eqref{eq:quotient}, and ignoring the background redshift, we obtain
\be
\label{eq:rfe}
1+\hat{z} = 1+ \left( \Ef\big|_{s}-\Ef\big|_{o} \right) - \Ef\big|_{s}\Ef\big|_{o} + \left(\Ef\big|_{o}\right)^{2} + \frac{1}{2}\left( \Es\big|_{s}-\Es\big|_{o} \right).
\ee
From \eq{eq:rfe} the redshift of a source $s$ is, at first order,
\be
\label{eq:dz1}
\dzf= \left( v_{1i}n^{i} + \ff\right)\big|^{o}_{s} + \int_{\lambda_{o}}^{\lambda_{s}}  \left[\ff' +\pf'\right]d\lambda,
\ee
where the integral runs w.r.t.~the affine parameter $\lambda$ along the line of sight. In \eq{eq:dz1} we identify the following elements
\begin{itemize}
\item[a)] \textit{The Doppler redshift}, which depends on the difference between the peculiar velocities of the source and the observer
\be
\dzf_{\text{Doppler}}=\left(v_{1 i}n^{i}\right)_{o} - \left(v_{1 i}n^{i}\right)_{s}.
\ee
\item[b)] \textit{The Gravitational redshift}, which describes the change of energy the photon experiences when it travels from a region with potential $\ff|_{s}$ to a region with potential $\ff|_{o}$
\be
\dzf_{\text{grav}}=\ff |_{o} - \ff|_{s}.
\ee
\item[c)] \textit{The Integrated Sachs-Wolfe effect (ISW)}, which describes the change in energy when a photon travels through a potential well between the source and the observer, this effect is only non-zero when the gravitational potential evolves during and along the photon's trajectory, so that the energy gained by going down the gravitational potential is not cancelled by the energy lost by climbing out the potential at the other end.
\be
\dzf_{\text{ISW}}=\int_{\lambda_{o}}^{\lambda_{s}}\left[\ff' +\pf'\right]d\lambda.
\ee
\end{itemize}

At second order the redshift \eqref{eq:dz2-metric} can also be decomposed as above
\begin{align}
\label{eq:dz2}
\dzs&= \dzs_{\text{Doppler}} + \dzs_{\text{grav}} + \dzs_{\text{ISW}} + \dzs_{\text{NL}},
\end{align}
where the nonlinear contribution from squared first order quantities denoted by `NL' is given by
\begin{align}
\dzs_{\text{NL}} &= \dzs_{\text{grav}\times\text{grav}} + \dzs_{\text{grav}\times\text{ISW}} + \dzs_{\text{IISW}}\\
&\quad + \dzs_{\text{grav}\times\text{Doppler}}+ \dzs_{\text{Doppler}\times\text{Doppler}}+ \dzs_{\text{Doppler}\times\text{ISW}}. \notag
\end{align}

Here, just as in the linear case, we have
\begin{itemize}
\item[(a)] Nonlinear Doppler redshift,
\begin{align}
\dzs_{\text{Doppler}} &= \frac{1}{2} \Big[\left(v_{2i}n^{i}\right)_{o} - \left(v_{2i}n^{i}\right)_{s} \Big] \notag,
\end{align}
\item[(b)] Nonlinear gravitational redshift,
\begin{align}
\dzs_{\text{grav}} &=  \frac{1}{2} \Big[ \fs|_{o} - \fs|_{s} \Big] , \notag 
\end{align}
\item[(c)] Nonlinear ISW,
\begin{align}
\dzs_{\text{ISW}} &= \frac{1}{2}\int_{\lambda_{o}}^{\lambda_{s}} \Big( \fs'+\ps' \Big) \dd\lambda .\notag
\end{align}
\end{itemize}

Here the second order contribution to the effects described above are evident, plus the product of first order contributions:
\begin{itemize}
\item[(1)] Gravitational redshift squared
\be
\dzs_{\text{grav} \times \text{grav}} = - \frac{3}{2} \left( \ff|_{s} - \ff|_{o}\right)^{2}+6 \ff|_{s}\ff|_{o}, \notag
\ee
\item[(2)] Gravitational redshift $\times$ ISW 
\be
\dzs_{\text{grav} \times \text{ISW}} = 2\left(\ff|_{s}-\ff|_{o}\right)\int_{\lambda_{o}}^{\lambda_{s}}\dd\lambda\left( \ff' + \pf' \right), \notag
\ee
\item[(3)] Gravitational redshift $\times$ Doppler
\be
\dzs_{\text{grav} \times \text{Doppler}} = \ff|_{o}\left( v_{1i}n^{i} \right)_{s} -\ff|_{s}\left( v_{1i}n^{i} \right)_{o} - 2 \pf|_{o}\left( v_{1i}n^{i} \right)_{o},\notag
\ee
\item[(4)] Doppler squared
\be
\dzs_{\text{Doppler} \times \text{Doppler}} = \frac{1}{2}\Big[\left( v_{1k}v^{k}_{1}\right)_{s}-\left( v_{1k}v^{k}_{1}\right)_{o}\Big] -\left( v_{1i}n^{i} \right)_{o}\Big[\left( v_{1i}n^{i}\right)_{s}-\left( v_{1i}n^{i}\right)_{o}\Big], \notag
\ee
\item[(5)] Doppler $\times$ ISW
\be
\dzs_{\text{Doppler} \times \text{ISW}} = \left( v_{1i}n^{i} \right)_{o}\int_{\lambda_{o}}^{\lambda_{s}}\dd\lambda\left( \ff' + \pf' \right) +  2v_{1 i}\int_{\lambda_{o}}^{\lambda_{s}}\dd \lambda\left( {\ff_{,}}^{i} + {\pf_{,}}^{i} \right), \notag
\ee
\item[(6)] Integrated ISW
\begin{align}
\dzs_{\text{IISW}} &= 2\int_{\lambda_{o}}^{\lambda_{s}}\dd\lambda\left[\pf' \left( \ff + \pf \right)\right] + 2\int_{\lambda_{o}}^{\lambda_{s}}\dd\lambda \left[ \ff \left( \frac{\dd \ff}{\dd \lambda} - 2\frac{\dd \pf}{\dd \lambda}\right)\right] \notag\\
& \quad -4 \int_{\lambda_{o}}^{\lambda_{s}}\dd\lambda\left[\left( \pf \ff_{,i} - \ff \pf_{,i} \right)n^{i}\right] + 4\int_{\lambda_{o}}^{\lambda_{s}}\dd\lambda\left[ n^{i}\ff_{,i}\left( \ff - \pf \right)\right]  \notag \\
& \quad + \int_{\lambda_{o}}^{\lambda_{s}}\dd\tilde{\lambda}\Bigg\{2 \left( \ff' - \pf'\right)\int_{\lambda_{o}}^{\lambda_{s}}\dd\lambda\left( \ff' + \pf' \right) \notag \\
&\quad- n^{i} \left( \ff_{,i} + \pf_{,i} \right) \int_{\lambda_{o}}^{\lambda_{s}}\dd\lambda\left( \ff' + \pf'\right) +  \left( \frac{\dd\ff}{\dd \lambda} - \frac{\dd \pf}{\dd \lambda}\right)\int_{\lambda_{o}}^{\lambda_{s}}\dd\lambda \left( \ff' + \pf' \right) \notag \\
& \quad +2\left( \ff-\pf\right)n^{i}\int_{\lambda_{o}}^{\lambda_{s}}\dd\lambda\left( \ff' + \pf'\right)_{,i} \notag \\
&\quad +\left( \int_{\lambda_{o}}^{\lambda_{s}}\dd\lambda{\left( \ff + \pf\right)_{,}}^{i}\right)\left( \int_{\lambda_{o}}^{\lambda_{s}}\dd\lambda\left( \ff' + \pf'\right)_{,i}\right)\notag \\
&\quad - n^{i}\left( \int_{\lambda_{o}}^{\lambda_{s}}\dd\lambda\left( \ff' + \pf' \right)_{,i} \right)\left( \int_{\lambda_{o}}^{\lambda_{s}}\dd\lambda\left( \ff' + \pf' \right) \right)\Bigg\}. \notag
\end{align}
\end{itemize}

\section{Distance determinations and the observed volume}
\label{sec:distances}

\subsection{Angular Diameter Distance}

To measure the angular diameter distance ($d_{A}$) we must define a projector into the screen space perpendicular to the light ray as shown in Fig. \ref{fig:da}. The screen space is orthogonal to the light ray {\textit{and}} to the observer 4-velocity. In fact the tensor 
\be
\P_{\mu \nu} = g_{\mu\nu}+ u_{\mu}u_{\nu}-n_{\mu}n_{\nu},
\ee
where $g_{\mu\nu}$ is the metric, $u_{\mu}$ is the 4-velocity and $n_{\mu}$ is the 4-direction of observation\footnote{Note that in the background, $n^{\mu}= a^{-1}[0,n^i]$.}, projects 4-vectors onto screen space, as can be seen in Fig.~\ref{fig:screen}. The projector tensor satisfies the relations
\be
{\P_{\mu}}^{\mu}=2, \qquad \P_{\mu\alpha}{\P^{\alpha}}_{\nu}=\P_{\mu\nu}, \qquad \P_{\mu\nu}k^{\nu} =\P_{\mu\nu}u^{\nu} = \P_{\mu\nu}n^{\nu} = 0.
\ee

The null expansion, $\theta$, and null shear, $\Sigma_{\mu \nu}$, are optical properties given in terms of the tangent vector $k^{\mu}$ by \cite{malik,sachs3,cc4}
\begin{align}
\label{eq:deftheta}
\theta &= \P^{\mu\nu}\nabla_{\mu}k_{\nu}, \\
 \label{eq:defsigma}
 \Sigma_{\mu \nu}& = {\P_{(\mu}}^{\sigma}{\P_{\nu)}}^{\rho}\nabla_{\sigma}k_{\rho}-\frac{1}{2} \theta \P_{\mu \nu},
\end{align}
Here $\theta$ describes the rate of expansion of the projected area of a bundle of light rays and $\Sigma_{\mu\nu}$ describes its rate of shear illustrated in Fig.~\ref{fig:da}.  Note that the wavevector can be obtained from a scalar potential ($S$), i.e.~$k_{\mu}=\nabla_{\mu}S$, and thus there is no null vorticity, that is $\omega \equiv \nabla_{[\mu}k_{\nu]} = 0$ \cite{cc7}.

The ``null evolution'' is given by the Sachs propagation equations (see e.g.~\cite{sachs3} for full derivation) 
\begin{align}
\label{eq:dth}
\frac{\dd \theta}{\dd \lambda} &= -\frac{1}{2} \theta^{2} - \Sigma_{\mu \nu}\Sigma^{\mu \nu} - R_{\mu \nu}k^{\mu}k^{\nu},\\
\label{eq:shr}
\frac{\dd \Sigma_{\mu \nu}}{\dd \lambda} &= -\Sigma_{\mu \nu} \theta + C_{\mu \rho \nu \sigma} k^{\rho} k^{\sigma},
\end{align}
where $C_{\mu \rho \nu \sigma}$ is the Weyl tensor. 

Eqs. \eqref{eq:dth} and \eqref{eq:shr} allow us to compute the angular diameter distance as a parametric function depending only on the affine parameter, $\lambda$, in contrast to previous works where the dependency is on the redshift  \cite{durrer2,cc1,cc2} or the conformal time \cite{yoo2}. The advantages of maintaining this dependency are discussed in section \ref{sec:discussion}.

From \eqs{eq:daa} and (\ref{eq:dth}) we obtain a second order differential equation for the angular diameter distance,
\begin{equation}
\label{eq:da}
\frac{\dd^{2}d_{A}}{\dd \lambda^{2}} = -\frac{1}{2} \left( R_{\mu \nu} k^{\mu}k^{\nu} +\Sigma_{\mu \nu} \Sigma^{\mu \nu} \right) d_{A}.
\end{equation}

We require appropriate initial conditions to solve \eqref{eq:da}. These can be found from the series expansion of the squared distance given by Kristian and Sachs in \cite{sachs2}:
\be
d_{A}^{2}(\lambda_{s}) = (u_{\mu}k^{\mu})_{o}^{2}(\lambda_{o}-\lambda_{s})\left[ 1 - \frac{1}{6}\left( R_{\mu\nu}k^{\mu}k^{\nu}\right)_{o}(\lambda_{o}-\lambda_{s})^{2}+\cdots \right],
\ee
from where we obtain the boundary conditions at the observer
\be
\label{eq:init}
d_{A}(\lambda_{o}) = 0, \quad \text{and}, \quad \frac{\dd d_{A}}{\dd \lambda}\Big|_{o} = -\E_{o}.
\ee

In this section we will define a conformal metric $g_{\mu \nu} = a^{-2} \hat{g}_{\mu \nu}$, useful to compute the angular diameter distance. In our notation, a hat ($ \hat{ \_ } $) denotes quantities on the physical spacetime, while quantities on the conformal spacetime have no hat. The background of the metric $g_{\mu \nu}$ is Minkowski spacetime, which simplifies both the equations and the calculations.
Conformal maps preserve both angles and shapes of infinitesimally small figures, but not their overall size \cite{conformal}. The conformal transformation $\hat{g}_{\mu\nu} \rightarrow g_{\mu\nu}$ maps the null geodesic equation of the perturbed FLRW metric $\hat{g}_{\mu \nu}$ to a null geodesic on the perturbed Minkowski metric $g_{\mu\nu}$ \cite{conformalnotes} and the angular diameter distance transforms as 
$
\hat{d}_{A}= a d_{A}.
$ 
The affine parameter transforms as 
$
\dd \hat{\lambda} = a^{2}\dd {\lambda}, 
$
so that the photon ray vector transforms as 
$
\hat{k}^{\mu} = a^{-2}k^{\mu} \iff \hat{k}_{\mu}=k_{\mu}
$ \cite{cc3}. For the 4-velocity we have 
$
\hat{u}_{\mu} = a u_{\mu}.
$
Finally, the energy transforms as 
$
\hat{\E}=-\hat{u}_{\nu}\hat{k}^{\nu} = -a^{-1}u_{\nu}k^{\nu}=a^{-1}\E.
$ 
In Minkowski spacetime we normalise $\E=1$. 

Hereafter, and until the end of this section, we will be working in a perturbed Minkowski spacetime, in order to finally conformally transform our result back to a FLRW spacetime.

In the Minkowski background, \eq{eq:da} simplifies to 
\be 
\frac{\dd^{2}\bar{d}_{A} }{\dd \lambda^{2}}= 0,
\ee
since $\bar{R}_{\mu \nu}$ and the shear vanish in the background. The solution is then 
\be
\bar{d}_{A}=C_{1}+\lambda C_{2}.
\ee 
The initial conditions given in \eq{eq:init} yield $C_{1}=0$ and $C_{2}=-1$, so that
\be
\label{eq:bgda-mink}
\bar{d}_{A}(\lambda_{s}) = \lambda_{o} - \lambda_{s}.
\ee
Mapping this into the FLRW background we obtain, for the angular diameter distance,
\be
\hat{d}_{A}(\hat{\lambda}_{s}) = a(\hat{\lambda}_{s})\left( \hat{\lambda}_{o}-\hat{\lambda}_{s}\right),
\ee
which can be expressed in terms of the comoving distance \eqref{eq:comovingd} as
\be
\label{eq:dacd}
\hat{d}_{A}(\hat{z}) = \frac{\chi(\hat{z})}{1+\hat{z}}.
\ee

\noindent Here we have used the definition of the scale factor $a(\hat{z})=1/(1+\hat{z})$ and the fact that the comoving distance depends on the redshift as given in Eq.~\eqref{eq:comd}. 

In general, at first order \eq{eq:da} takes the form
\be
\label{eq:da2}
\frac{\dd^{2} \daf}{\dd \lambda^{2}} = -\frac{1}{2}\left[ 2 \bar{R}_{\mu \nu} \bar{k}^{\mu} \dkf^{\nu} \bar{d}_{A} + \dRf_{\mu \nu} \bar{k}^{\mu} \bar{k}^{\nu} \bar{d}_{A} + \bar{R}_{\mu \nu} \bar{k}^{\mu} \bar{k}^{\nu} \daf\right] - \bar{d}'_{A} \frac{\dd\dnuf}{\dd \lambda}-2 \bar{d}''_{A} \dnuf.
\ee
where we use take the background equivalence between the affine parameter and the conformal time $\dd \lambda = \dd \eta$, so that
\be
\frac{\dd \bar{d}_{A}}{\dd \eta} = \frac{\dd \bar{d}_{A}}{\dd \lambda}.
\ee 

\noindent This relation is only fulfilled in the background. Once perturbations are introduced the relation between the affine parameter and time becomes non-trivial.

In Minkowski spacetime, \eq{eq:da2} simplifies to
\begin{align}
\label{eq:dda1}
\frac{\dd^{2} \daf}{\dd \lambda^{2}} &= -\frac{1}{2}\bar{d}_{A}\left[2 \left( \frac{\dd^{2}\pf}{\dd\lambda^{2}}\right) +\nabla^{2}\left( \ff + \pf \right)-n^{i}n^{j}\left( \ff + \pf\right)_{,ij}-\frac{2}{\bar{d}_{A}}\frac{\dd \dnuf}{\dd \lambda}\right],
\end{align}
where we used the background solution for $d_{A}$ \eqref{eq:bgda-mink}, the first order perturbation of the Ricci tensor $\dRf_{\mu\nu}$ given in Appendix \ref{riccipert}, and \eqs{eq:der-affine} and \eqref{eq:double-der-affine}.

The solution to \eqref{eq:dda1} is, upon several integrations by parts, 
\begin{align}
\label{eq:da-linear}
\frac{\daf(\lambda_{s})}{\bar{d}_{A}(\lambda_{s})} &=   \ff|_{o}-\pf|_{o} - \pf|_{s} - \left(v_{1i}n^{i} \right)_{o} - \frac{1}{\lambda_{o}-\lambda_{s}}\Bigg\{2\int_{\lambda_{o}}^{\lambda_{s}}\dd\lambda \pf \\
& \quad +\frac{1}{2} \int_{\lambda_{o}}^{\lambda_{s}}\dd\lambda\left(\lambda_{s}-\lambda\right)\left( \lambda_{o}-\lambda\right)\left[ \nabla^{2}\left( \ff+\pf \right) - n^{i}n^{j}\left( \ff+\pf \right)_{,ij} \right.\notag \\
& \left. \qquad \qquad \qquad \qquad \qquad \qquad \qquad \qquad \qquad \qquad \qquad \qquad -\frac{2}{\lambda_{o}-\lambda_{s}}\frac{\dd\dnuf}{\dd \lambda} \right]\Bigg\}. \notag
\end{align}
From \eq{eq:init}, in general
\be
\label{eq:da-linear-no-slip}
\delta^{(n)} d_{A}(\lambda_{o}) = 0, \quad \text{and}, \quad \frac{\dd \delta^{(n)}d_{A}}{\dd \lambda}\Big|_{o} = - \delta^{(n)}\E_{o}.
\ee
In the absence of anisotropic stress, $\ff = \pf$ (see, e.g.~Ref.~\cite{malik}), we recover in \eq{eq:da-linear} the fully relativistic lensing convergence, usually denoted as $\kappa$ \cite{durrer1,cc5,cc7,yoo2}, at first order, which includes Sachs-Wolfe (SW), Integrated Sachs-Wolfe (ISW) and Doppler terms in addition to the standard lensing integral

\begin{align}
\label{eq:da-noan}
\frac{\daf(\lambda_{s})}{\bar{d}_{A}(\lambda_{s})} &=  - \ff|_{s} - \left(v_{1i}n^{i} \right)_{o} - \frac{1}{\lambda_{o}-\lambda_{s}}\Bigg\{2\int_{\lambda_{o}}^{\lambda_{s}}\dd\lambda \ff \\
& \quad + \int_{\lambda_{o}}^{\lambda_{s}}\dd\lambda\left(\lambda_{s}-\lambda\right)\left( \lambda_{o}-\lambda\right)\left[ \nabla^{2}\left( \ff \right) - n^{i}n^{j}\left( \ff \right)_{,ij} -\frac{1}{\lambda_{o}-\lambda_{s}}\frac{\dd\dnuf}{\dd \lambda} \right]\Bigg\}. \notag
\end{align}

At second order, \eq{eq:da} takes the form
\begin{align}
\label{eq:da-two}
\frac{\dd^{2} \das}{\dd \lambda^{2}} &=  -\Bigg[ 2\bar{k}^{\mu}\dkf^{\nu} \dRf_{\mu\nu} +\frac{1}{2} \bar{k}^{\mu}\bar{k}^{\nu}\dRs_{\mu\nu}+\dS_{\mu\nu}\dS^{\mu\nu} \Bigg]\bar{d}_{A} \\
&\quad - \Big[ \bar{k}^{\mu}\bar{k}^{\nu}\dRf_{\mu\nu} \Big]\daf -2 \Big[ \dkf^{\mu}\nabla_{\mu}\dnuf + 3 \dkf^{\mu} \bar{k}^{\alpha}\Gamma_{\mu\alpha}^{0} \Big]\bar{d}_{A}' \notag \\
&\quad - 2\Big[ \dnus + \left( \dnuf \right)^{2} \Big]\bar{d}_{A}'' - \left( \frac{\dd \dnus}{\dd \lambda} \right)\bar{d}_{A}' -4 \left(\dnuf\right)\left( \daf ''\right) \notag \\
&\quad -2\Bigg[ \bar{k}^{\mu}\bar{k}^{\alpha}\Gamma_{\mu\alpha} + \frac{\dd \dnuf}{\dd \lambda} \Bigg]\daf' \notag,
\end{align}
where $\dS_{\mu \nu}$ is the linear perturbation to the shear. Using \eq{eq:defsigma} we obtain
\be
\label{eq:ddS}
\frac{\dd \dS_{ij}}{\dd \lambda} = \frac{1}{2}\delta_{ij}\nabla^{2}\left(\ff + \pf\right) - \left( \ff + \pf\right)_{,ij}. 
\ee
Without loss of generality, we set the perturbation of the shear at the observer $\dS^{\mu\nu}|_{o} = 0$, and integrating along the line of sight from the observer to the source ($\lambda_{o}$ to $\lambda_{s}$), we obtain
\be
\label{eq:dSij}
\dS_{ij} = \int_{\lambda_{o}}^{\lambda_{s}}\dd \lambda \Bigg[ \frac{1}{2}\delta_{ij}\nabla^{2}\left(\ff + \pf\right) - \left( \ff + \pf\right)_{,ij} \Bigg] . 
\ee
The contraction $\dS_{ij}\dS^{ij}$ is given by
\begin{align}
\label{eq:contracted-null-shear}
\dS_{ij} \dS^{ij} &= \Bigg[  \int_{\lambda_{o}}^{\lambda_{s}}\dd \lambda \Bigg[ \frac{1}{2}\delta_{ij}\nabla^{2}\left(\ff + \pf\right) - \left( \ff + \pf\right)_{,ij} \Bigg] \Bigg]\times \\ 
&\qquad\qquad\qquad \Bigg[  \int_{\lambda_{o}}^{\lambda_{s}}\dd \lambda \Bigg[ \frac{1}{2}\delta^{ij}\nabla^{2}{\left(\ff + \pf\right) - \left( \ff + \pf\right)_{,}}^{ij} \Bigg] \Bigg] , \notag\\
&= \left(\int_{\lambda_{o}}^{\lambda_{s}}\dd\lambda\left( \ff + \pf \right)_{,ij}\right)\left( \int_{\lambda_{o}}^{\lambda_{s}}\dd\lambda{\left( \ff + \pf\right)_{,}}^{ij}\right) -\frac{1}{4}\left[ \int_{\lambda_{o}}^{\lambda_{s}}\dd\lambda\nabla^{2}\left( \ff + \pf \right)\right]^{2}. \notag
\end{align}

In Eq.~\eqref{eq:da-second-metric} we find the second order part of the angular diameter distance, using the background solution for $\bar{d}_{A}$, and the full expression $\dRs_{\mu \nu}$, the second order perturbation of the Ricci tensor given in Appendix \ref{riccipert}.
 
Thus, the total area distance as a function of the affine parameter in a perturbed FLRW spacetime is given by
\be
\label{eq:daaff}
\hat{d}_{A}(\lambda_{s}) = a(\lambda_{s})(\lambda_{o}-\lambda_{s})\left[ 1 + \frac{\daf(\lambda_{s})}{\bar{d}_{A}(\lambda_{s})}+\frac{1}{2}\frac{\das(\lambda_{s})}{\bar{d}_{A}(\lambda_{s})} \right],
\ee
where the solutions for $\bar{d}_{A}(\lambda_{s})$, $\daf(\lambda_{s})$ and $\das(\lambda_{s})$ are given in \eqs{eq:bgda-mink}, \eqref{eq:da-linear} and \eqref{eq:da-second-metric}, respectively. From here onwards, we abandon the conformal Minkowski spacetime and return to FLRW.

\subsection{Area distance as a function of observed redshift}
\label{dafredshift}

In order to compare with previous work done in the literature, we can convert the angular diameter distance in terms of the affine parameter to a function of the observed redshift. To do so, we need to perturbatively invert $z(\lambda)$ into $\lambda(z)$ and substitute this into Eq.~\eqref{eq:daaff}. This means we need $d_{A}$ on surfaces of constant observed redshift $z$ rather than on surfaces of constant affine parameter $\lambda$, which is not observable.

We expand the affine parameter in perturbation theory as 
\be
\label{eq:pertl}
\lambda = \varsigma + \delta^{(1)} \lambda + \frac{1}{2} \delta^{(2)} \lambda,
\ee
where $\varsigma$ is the affine parameter in redshift space corresponding to the redshift $\hat{z}$, \textit{as if there were no perturbations} \cite{cc2}. We define $\varsigma$ using as an anchor the background relation
\be
\label{eq:anu}
a(\varsigma) = \frac{1}{1+\hat{z}},
\ee
with this relation we can fix $\delta^{(1)}\lambda$ and $\delta^{(2)}\lambda$, since it should always hold, and if there are any perturbations, they should cancel since Eq.~\eqref{eq:anu} is only valid in the background. To begin with, we see that at any redshift $\hat{z}$, the derivatives of $a$ with respect to $\varsigma$ are
\begin{align}
\frac{1}{a} \frac{\dd a}{\dd \varsigma} &= \H(\varsigma), \\
\frac{1}{a} \frac{\dd^{2} a}{\dd \varsigma^{2}} &= \left[ \frac{\dd \H(\varsigma)}{\dd \varsigma} + \H^{2}(\varsigma) \right]. 
\end{align}
We now expand the scale factor $a$ about $\lambda$ up to and including second order perturbations as defined in Eq.~\eqref{eq:pertl}, we have
\be
\label{eq:anutaylor}
a(\lambda) = a(\varsigma) \left[ 1 +\H \delta^{(1)} \lambda + \frac{1}{2}\H \delta^{(2)} \lambda +\frac{1}{2} \left( \frac{\dd \H}{\dd \lambda} +\H^{2} \right)\left(\delta^{(1)}\lambda\right)^{2} + \mathcal{O}\left(\delta^{(3)}\lambda\right) \right].
\ee
Using Eqs.~\eqref{eq:redshift1} and \eqref{eq:anutaylor} we find that, 
\begin{align}
\label{eq:tayloranu}
\frac{1}{1+\hat{z}} &= \frac{a(\varsigma)}{a(\varsigma_{o})}\Bigg[ 1+ \left( \H \delta^{(1)} \lambda - \dzf \right)   \\
& \qquad +\frac{1}{2}\left\{ \H \delta^{(2)}\lambda - \dzs + 2\left( \dzf\right)^{2} - 2\H \delta^{(1)}\lambda \dzf + \left( \frac{\dd \H}{\dd \varsigma} + \H^{2} \right)\left(\delta^{(1)}\lambda\right)^{2}\right\} \Bigg]. \notag
\end{align}
From the background relation given in Eq.~\eqref{eq:anu} and \eqref{eq:tayloranu} we then find that the perturbations to the affine parameter must follow the following relations:
\begin{align}
\label{eq:pertlz1}
\delta^{(1)}\lambda &= \frac{\dzf}{\H}, \\
\label{eq:pertlz2}
\delta^{(2)}\lambda &= \frac{1}{\H}\left[ \dzs - \left( \dzf \right)^{2}\left( 1 + \frac{1}{\H^{2}}\frac{\dd \H}{\dd \varsigma} \right) \right].
\end{align}
Finally, using these relations to substitute for $a(\lambda_{s})\left( \lambda_{o}-\lambda_{s}\right)$, we find that the area distance \eqref{eq:daaff} becomes
\begin{align}
\label{eq:danu1}
\hat{d}_{A} (\varsigma)&= a(\varsigma)\left( \varsigma_{o}-\varsigma\right)\Bigg\{ 1 + \left[ \frac{\daf}{\bar{d}_{A}}(\lambda)+\left( 1-\frac{1}{\varsigma_{o}-\varsigma}\dzf(\lambda)\right) \right] \\
& \quad + \frac{1}{2}\left[ \frac{\das}{\bar{d}_{A}}(\lambda) +\left( 1-\frac{1}{\varsigma_{o}-\varsigma}\dzs(\lambda)\right)  + \frac{\H'-\H^{2}}{\H^{3}(\varsigma_{o}-\varsigma)}\left( \dzf \right)^{2} \right. \notag \\
& \quad \left. +2\left( 1 - \frac{1}{\H(\varsigma_{o}-\varsigma)}\right) \frac{\daf}{\bar{d}_{A}}\dzf \right] \Bigg\}. \notag
\end{align}
Up until here we have corrected the scale factor from the affine parameter $\lambda$ to $\varsigma$. Now we need to convert the first order contributions because they bring additional second order contributions. We introduce that, for a general first order quantity $\delta^{(1)}X$, converting to $\varsigma$ gives
\be
\delta^{(1)}X(\lambda) = \delta^{(1)}X(\varsigma) + \frac{\partial \delta^{(1)} X}{\partial \lambda}\Big|_{\varsigma} \frac{\dzf (\varsigma)}{\H},
\ee
where $\delta^{(1)}X(\varsigma)$ is to be understood as substituting $\varsigma$ in the expression for $\delta^{(1)}X(\lambda)$, i.e. $\delta^{(1)} X(\lambda \to \varsigma)$. The factor $\partial_{\lambda}\delta^{(1)}X|_{\varsigma}$ is multiplied by a first order quantity, so the derivative is evaluated in the background. Thus, we can write $\dd_{\varsigma}\delta^{(1)}X$.
With this, Eq.~\eqref{eq:danu1} finally becomes
\begin{align}
\label{eq:da-sigma}
\hat{d}_{A}(\hat{z}) &= \frac{\varsigma_{o}-\varsigma}{1+\hat{z}} \Bigg\{ 1 + \left[ \frac{\daf}{\bar{d}_{A}} + \left( 1-\frac{1}{\H(\varsigma_{o}-\varsigma)} \right)\dzf \right] \\
&\quad + \frac{1}{2}\left[ \frac{\das}{\bar{d}_{A}} + \left( 1-\frac{1}{\H(\varsigma_{o}-\varsigma)} \right)\dzs  \right. \notag \\
&\quad  +2\left( 1-\frac{1}{\H(\varsigma_{o}-\varsigma)} \right) \left( \frac{\daf}{\bar{d}_{A}} + \frac{1}{\H}\frac{\dd \dzf}{\dd \varsigma} \right)\dzf \notag \\
&\quad\left.  + 2\frac{\dd}{\dd \varsigma}\left(\frac{\daf}{\bar{d}_{A}}\right) \frac{\dzf}{\H} +\frac{\H'-\H^{2}}{\H^{3}(\varsigma_{o}-\varsigma)} \left( \dzf \right)^{2} \right]\Bigg\}. \notag
\end{align}

\noindent We can now write the angular diameter distance as a function of the redshift, although it is written in terms of integrals over the comoving distance $\chi = \varsigma_{o}-\varsigma$, which depends on the redshift itself by Eq.~\eqref{eq:comd}.

Using Eq.~\eqref{eq:da-sigma} written in terms of observable quantities such as the observable redshift and comoving distance, the angular diameter distance will possibly, in principle, be measured with great accuracy by the upcoming surveys, and should complement to the known luminosity distance measurements quite well.

Combining Eqs.~\eqref{eq:dz1}, \eqref{eq:da-noan} with \eqref{eq:da-sigma}, we have that at linear order the diameter distance as a function of redshift is given by,
\begin{align}
\label{eq:dazf}
\delta^{(1)}\hat{d}_{A}(\hat{z}_{s}) &= \frac{\chi_{s}}{1+\hat{z}_{s}}\Bigg\{ -\pf|_{s} - \pf|_{o} - \left( 1 - \frac{1}{\H \chi_{s}}\right)\ff|_{s} + \left( 2-\frac{2}{\H \chi_{s}} \right)\left( v_{1i}n^{i}\right)_{o} \notag \\
& \quad\quad + \left( 1-\frac{1}{\H \chi_{s}}\right)\left( v_{1i}n^{i}\right)_{s} + \left( 1 - \frac{1}{\H\chi_{s}}\right)\int_{0}^{\chi_{s}}\left( \ff'+\pf'\right)\dd\chi -\frac{2}{\chi_{s}}\int_{0}^{\chi_{s}}\pf\dd\chi \notag \\
& \quad \quad -\frac{1}{2 \chi_{s}} \int_{0}^{\chi_{s}}\dd\chi \left( \chi - \chi_{s} \right)\chi \left[ \nabla^{2}\left( \ff + \pf\right) - n^{i}n^{j}\left( \ff + \pf \right)_{,ij} - \frac{2}{\chi} \frac{\dd \dnuf}{\dd \varsigma} \right] \Bigg\}. 
\end{align}

The full expression for $\delta^{(2)}\hat{d}_{A}(\hat{z}_{s})$ in terms of the metric potentials is given in Appendix \ref{metriccomp}.

\subsection{Physical Volume}

The area distance the light-ray bundle creates, changes along the line of sight as seen in Fig.~\ref{fig:screen}, and we are interested in computing the volume that these hypersurfaces enclose, since therein lie the overdensities we are accounting for.

The volume element \eqref{eq:dldo} can be rewritten in terms of the quantities we have computed in the previous sections; the angular diameter distance in \eqs{eq:da-noan} and \eqref{eq:da-second-metric}, and the energy in \eqs{eq:dE1} and \eqref{eq:dE2}. It is given up to second order by,
\begin{align}
\label{eq:volume}
\dd V &= -\E d_{A}^{2}(\lambda) \dd \lambda \dd \Omega,  \\
&= - \bar{\E}\bar{d}_{A}^{2} \Bigg[ 1+ 2 \frac{\daf}{\bar{d}_{A}}+\frac{\Ef}{\bar{\E}} \notag \\ 
&\qquad\qquad\qquad\qquad+ \left(\frac{\daf}{\bar{d}_{A}}\right)^{2} + \left( \frac{\Ef}{\bar{\E}}\right) \left( \frac{\daf}{\bar{d}_{A}}\right) +\frac{\das}{\bar{d}_{A}} +\frac{1}{2}\frac{\Es}{\bar{\E}} \Bigg]\dd \lambda \dd \Omega. \notag
\end{align}
The volume element is given in terms of the affine parameter $\lambda$, but we need to express our result in terms of the observed redshift $z$, and so we need to take the volume in bins of $\dd z$ instead of $\dd \lambda$. To do so we use the fact that we can write the affine parameter as a function of redshift, i.e.~$\lambda(z)$, and using \eqs{eq:pertl}, \eqref{eq:pertlz1} and \eqref{eq:pertlz2}, we obtain the relation
\begin{align}
\frac{\dd\lambda}{\dd z} &= -\frac{a}{\H}\left[ 1 + \left(\frac{1}{\H}+\frac{1}{\H\left( 1+\bar{z}\right)}\right) \frac{\dd \dzf}{\dd \lambda} - \frac{\H'}{\H^{2}} \dzf\right] \\
&\qquad -\frac{a}{2\H}\Bigg[ \left(\frac{1}{\H}+\frac{1}{\H\left( 1+\bar{z}\right)}\right) \frac{\dd\dzs}{\dd\lambda} -\frac{\H'}{\H^{2}}\dzs -\frac{\H'}{\H^{3}\left( 1+\bar{z} \right)}\left( \frac{\dd\dzf}{\dd\lambda}\right)\dzf\notag \\
&\qquad +\frac{1}{\H^{2}\left( 1+\bar{z}\right)}\left( 2+\frac{1}{1+\bar{z}}  \right)\left( \frac{\dd\dzf}{\dd\lambda} \right)^{2} + \left( \frac{\H'}{\H^{2}} \right)\left( 1+\frac{\H'}{\H^{2}} \right)\left( \dzf \right)^{2} \notag \\
&\qquad -\frac{1}{\H} \left( 2 \frac{\left( \H' \right)^{2}}{\H^{2}} - \frac{\H''}{\H}\right)\left( \dzf \right)^{2} - \frac{1}{\H}\left( 1+\frac{\H'}{\H^{2}}\right)\frac{\dd}{\dd \lambda}\left[\left( \dzf \right)^{2}\right] \Bigg],\notag 
\end{align}
modifying Eq.~\eqref{eq:volume} into
\begin{align}
\label{eq:volumedz}
\dd V(z) &= -\E(z) d_{A}^{2}(z) \left(\frac{\dd \lambda}{\dd z}\right) \dd z \dd \Omega, \notag \\
&= \frac{\bar{\E}\bar{d}_{A}^{2}}{\H (1+z)} \Bigg[ 1+ \frac{2}{\H}\frac{\dd \dzf}{\dd \lambda} - \frac{\H'}{\H^{2}}\dzf +2 \frac{\daf}{\bar{d}_{A}}+\frac{\Ef}{\bar{\E}} \\
&\qquad - \frac{1}{\H} \frac{\dd\dzs}{\dd\lambda} +\frac{1}{2}\frac{\H'}{\H^{2}}\dzs -\frac{3}{2\H^{2}}\left( \frac{\dd\dzf}{\dd\lambda} \right)^{2} - \frac{1}{2}\left( \frac{\H'}{\H^{2}} \right)\left( 1+\frac{\H'}{\H^{2}} \right)\left( \dzf \right)^{2} \notag \\
&\qquad +\frac{1}{2\H^{2}} \left( 2 \frac{\left( \H' \right)^{2}}{\H^{2}} - \frac{\H''}{\H}\right)\left( \dzf \right)^{2} + \frac{1}{2\H}\left( 1+2\frac{\H'}{\H^{2}}\right)\left( \frac{\dd\dzf}{\dd\lambda}\right)\dzf\notag  \\
& \qquad  + \left(\frac{\daf}{\bar{d}_{A}}\right)^{2} + \left( \frac{\Ef}{\bar{\E}}\right) \left( \frac{\daf}{\bar{d}_{A}}\right) +\frac{\das}{\bar{d}_{A}} +\frac{1}{2}\frac{\Es}{\bar{\E}} \Bigg]\dd z \dd \Omega. \notag
\end{align}

We now give an expression for the volume element order by order. In the background we have
\begin{align}
\label{eq:bgV}
\dd\bar{V} &= - \bar{\E} \bar{d}_{A}^{2} \dd \lambda \dd \Omega = a^{2}\left(\lambda_{s}\right) [\lambda_{s}-\lambda_{o}]^{2}\dd \lambda \dd \Omega, \\
&= \frac{\bar{\E}\bar{d}_{A}^{2}}{\H(1+z)} \dd z \dd \Omega= \frac{\chi^{2}}{\H(1+z)}.
\end{align}

From \eq{eq:volumedz} and using \eqs{eq:dE1-metric} and \eqref{eq:dazf}, we have that the first order perturbation to the physical volume is
\begin{align}
\label{eq:dvv1}
\dd \dVf &= \frac{\bar{\E} \bar{d}^{2}_{A}}{\H(1+z)}\Bigg[ \frac{2}{\H}\frac{\dd \dzf}{\dd \varsigma} - \frac{\H'}{\H^{2}}\dzf + 2 \frac{\daf}{\bar{d}_{A}} + \frac{\delta^{(1)}\E}{\bar{\E}} \Bigg] \dd z \dd \Omega, \notag \\
&= \frac{\chi^{2}}{\H(1+z)}\Bigg[ \frac{2}{\H}\left( \pf' - \partial_{\chi}\ff + \frac{\dd \left( v_{1i}n^{i}\right)}{\dd \varsigma}\right) - 2\left( \ff + \pf \right) -3 \left( v_{1i}n^{i}\right) \notag \\
&\quad +\left( \frac{\H'}{\H^{2}} + \frac{2}{\H \chi} \right)\left( \ff - \left( v_{1i}n^{i} \right) + \int_{0}^{\chi}\dd \tilde{\chi}\left( \ff' + \pf'\right) \right) - \frac{4}{\chi}\int_{0}^{\chi}\dd\tilde{\chi}\pf \notag \\ 
&\quad  -\frac{1}{\chi}\int_{0}^{\chi}\dd \tilde{\chi}\left( \tilde{\chi}-\chi \right)\tilde{\chi} \left\{ \nabla^{2}\left( \ff + \pf \right) - n^{i}n^{j}\left( \ff + \pf\right)_{,ij} - \frac{2}{\tilde{\chi}} \frac{\dd \dnuf}{\dd \varsigma} \right\} \notag \\ 
&\quad + 3\int_{0}^{\chi}\dd \tilde{\chi}\left( \ff' + \pf' \right)\Bigg],
\end{align}
and using \eqs{eq:dE2-metric} and \eqref{eq:da-second-metric} we find that the second order perturbation to the physical volume is 
\begin{align}
\label{eq:dvv2}
\dd \dVs &= \frac{\bar{\E}\bar{d}_{A}^{2}}{\H (1+z)} \Bigg[- \frac{1}{\H} \frac{\dd\dzs}{\dd\lambda} +\frac{1}{2}\frac{\H'}{\H^{2}}\dzs -\frac{3}{2\H^{2}}\left( \frac{\dd\dzf}{\dd\lambda} \right)^{2} - \frac{1}{2}\left( \frac{\H'}{\H^{2}} \right)\left( 1+\frac{\H'}{\H^{2}} \right)\left( \dzf \right)^{2} \notag \\
&\qquad +\frac{1}{2\H^{2}} \left( 2 \frac{\left( \H' \right)^{2}}{\H^{2}} - \frac{\H''}{\H}\right)\left( \dzf \right)^{2} + \frac{1}{2\H}\left( 1+2\frac{\H'}{\H^{2}}\right)\left( \frac{\dd\dzf}{\dd\lambda}\right)\dzf\notag  \\
& \qquad  + \left(\frac{\daf}{\bar{d}_{A}}\right)^{2} + \left( \frac{\Ef}{\bar{\E}}\right) \left( \frac{\daf}{\bar{d}_{A}}\right) +\frac{\das}{\bar{d}_{A}} +\frac{1}{2}\frac{\Es}{\bar{\E}} \Bigg]\dd z \dd \Omega. 
\end{align}

\noindent The equivalent expression in terms of the metric potentials is given in Appendix \ref{metriccomp}. With the above expansion at hand, we have all the necessary quantities to compute, in the next section, the galaxy number density up to second order, our main result.

\section{Galaxy number density}
\label{sec:counts}
%
%

In this section we present our main result, the galaxy number overdensity at second order. As a first element, we take $V(n^{i},z)$ as the physical survey volume density per redshift bin per solid angle given by (\ref{eq:dldo}), where $n^{i}$ is the direction of observation and $z=z(\lambda_{s})$. The volume is a perturbed quantity since the solid angle of observation as well as the redshift bin are distorted between the source and the observer
\be
V(n^{i},z) = \bar{V}(z)+\dVf(n^{i},z)+\frac{1}{2}\dVs(n^{i},z).
\ee 
In \eqs{eq:dvv1} and \eqref{eq:dvv2} we provide the first and second order perturbations to the volume, respectively. Note that we use $\delta (\dd  V) /\dd \bar{V}$ where other authors in the literature use $\delta V / \bar{V}$ (see, e.g. Ref~\cite{durrer1}).

In a galaxy redshift survey, we measure the number of galaxies in direction $n^{i}$ at redshift $z$. Let us call this $N(n^{i},z) \dd \Omega_{n}\dd z$, where $\dd \Omega_{n}$ is the solid angle the survey spans. Then one must average over the angles to obtain their redshift distribution, $\langle N \rangle(z)\dd z$, where the angle brackets correspond to this angular average \cite{campos}
\be
\langle N\rangle(z) \dd z = \dd z \int_{\Omega_{n}} N(n^{i},z)\dd \Omega,
\ee
where the integral is over the solid angle the survey spans.

We can then build the matter density perturbation, density contrast, in redshift space, i.e. the perturbation variable \cite{durrer1}
\be
\delta_{z}(n^{i}, z) \equiv \frac{\rho(n^{i},z)-\langle \rho \rangle(z)}{\langle \rho \rangle(z)}.
\ee
and expand it up to second order as
\begin{align}
\delta_{z}(n^{i}, z) = \delta^{(1)}_{z}(n^{i},z) + \frac{1}{2}\delta^{(2)}_{z}(n^{i},z).
\end{align}

Our aim in the following is to compute the observed matter density perturbation since the density of sources is proportional to the number of the sources within a given volume, i.e.
\be
\label{eq:rho}
\rho(n^{i},z) = \frac{N(n^{i},z)}{V(n^{i},z)},
\ee
and expanding \eq{eq:rho} we show that at any order
\begin{equation}
\delta_{z}(n^{i}, z) = \frac{N(n^{i},z)-\langle N \rangle(z)}{\langle N \rangle(z)} - \frac{\delta V(n^{i}, z)}{V(z)}.
\end{equation}

The observed quantity is the perturbation in the number density of galaxies, $\Delta$, and it is defined as
\begin{equation}
\label{eq:ng}
\Delta(n^{i},z) \equiv \frac{N(n^{i},z)-\langle N \rangle(z)}{\langle N \rangle(z)} =\delta_{z}(n^{i}, z) + \frac{\delta V(n^{i}, z)}{V(z)},
\end{equation}
and we thus have
\begin{align}
\label{eq:defdg1}
\Delta_{g}^{(1)}(n^{i},z) &= \delta_{z}^{(1)}(n^{i},z) + \frac{\dVf(n^{i},z)}{\bar{V}(z)}, \\
\label{eq:defdg2}
\Delta_{g}^{(2)} (n^{i},z)&= \delta_{z}^{(2)}(n^{i},z) + \frac{\dVs(n^{i},z)}{\bar{V}(z)} + \delta_{z}^{(1)}(n^{i},z)\frac{\dVf(n^{i},z)}{\bar{V}(z)}.
\end{align}

In order to compute the above, let us first relate $\delta_{z}(n^{i},z)$ to the matter density quantity $\delta(x^{i},\eta)$ and the perturbations on the redshift computed in Section \ref{sec:whatwemeasure}. The redshift density up to second order in redshift space is
\begin{align}
\label{eq:redshift-density}
\delta_{z}(n^{i},z) &= \frac{\rho(n^{i},z)-\bar{\rho}(z)}{\bar{\rho}(z)} = \frac{\bar{\rho}(z)+\drf(n^{i},z)+\frac{1}{2}\drs(n^{i},z)-\bar{\rho}(z)}{\bar{\rho}(z)} \\
&= \frac{\bar{\rho}(\bar{z}+\dzf+\frac{1}{2}\dzs)+\drf(n^{i},z)+\frac{1}{2}\drs(n^{i},z)-\bar{\rho}(z)}{\bar{\rho}(z)} \notag\\
&= \frac{\drf(n^{i},z)}{\bar{\rho}(z)} + \frac{\dd \bar{\rho}}{\dd \bar{z}} \frac{\dzf(n^{i},z)}{\bar{\rho}(\bar{z})} \notag \\
&\qquad +\frac{1}{2}\frac{\drs(n^{i},z)}{\bar{\rho}(z)} + \frac{1}{2}\frac{\dd \bar{\rho}}{\dd \bar{z}} \frac{\dzs(n^{i},z)}{\bar{\rho}(\bar{z})}+\frac{1}{2}\frac{\dd^{2} \bar{\rho}}{\dd \bar{z}^{2}} \frac{\left[\dzf(n^{i},z)\right]^{2}}{\bar{\rho}(\bar{z})} + \frac{\dd \drf}{\dd \bar{z}}\frac{\dzf(n^{i},z)}{\bar{\rho}(\bar{z})}.\notag 
\end{align}
Structure in the universe is formed from dark matter and baryons which at large scales are modelled by a single pressureless component, which evolves with redshift as
\be
\label{eq:bg-rho}
\bar{\rho}(z)\approx \rho_{0}(1+z)^{3}.
\ee
Thus we have that
\be
\label{eq:der1-rho}
\frac{\dd \bar{\rho}}{\dd \bar{z}} = 3 \frac{\bar{\rho}}{1+\bar{z}},
\ee
so using \eq{eq:dz1}, the redshift density perturbation at first order is given by
\be
\label{eq:ddz1}
\delta^{(1)}_{z}(n^{i},z) = \frac{\drf(n^{i},z)}{\bar{\rho}(z)} + \frac{3}{1+\bar{z}}\left[\left( v_{1i}n^{i} + \ff\right)\big|^{s}_{o} + \int_{0}^{\chi_{s}}\dd\chi  \left\{\ff' +\pf'\right\}\right].
\ee

Combining (\ref{eq:dvv1})
and (\ref{eq:ddz1}) we find that the galaxy number density fluctuation in redshift space as defined in Eq.~(\ref{eq:defdg1}) is, at first order, 
\begin{align}
\label{eq:Dg1}
\Delta^{(1)}_{g}(n^{i},z) &= \left[\frac{\drf(n^{i},z)}{\bar{\rho}(z)} + 3\ff\right] -2\left(\ff + \pf\right) + \frac{1}{\H}\left( \pf' - \partial_{\chi}\ff + \frac{\dd \left(v_{1i}n^{i}\right)}{\dd \varsigma} \right) \\
& \quad +\left( \frac{\H'}{\H^{2}} +\frac{2}{\H \chi}\right)\left[ \ff - \left( v_{1i}n^{i}\right) + \int_{0}^{\chi}\dd\tilde{\chi}\left( \ff' + \pf' \right)\right] - \frac{4}{\chi}\int_{0}^{\chi}\dd \tilde{\chi}\pf\notag \\
& \quad - \frac{1}{\chi}\int_{0}^{\chi}\dd \tilde{\chi}\left( \tilde{\chi}-\chi\right)\tilde{\chi}\left[ \nabla^{2}\left( \ff + \pf\right) - n^{i}n^{j}\left( \ff + \pf \right)_{,ij} - \frac{2}{\tilde{\chi}}\frac{\dd \dnuf}{\dd \varsigma} \right]. \notag
\end{align}
From \eq{eq:bg-rho}, we have that the second derivative of the background density is
\be
\label{eq:der2-rho}
\frac{\dd^{2} \bar{\rho}}{\dd \bar{z}^{2}} = 6 \frac{\bar{\rho}}{(1+\bar{z})^{2}},
\ee  
so using \eqs{eq:dz1}, \eqref{eq:dz2} and \eqref{eq:der2-rho} in \eq{eq:redshift-density}  we find that the redshift density perturbation at second order is given by
\begin{align}
\delta^{(2)}_{z}(n^{i},z) &= \frac{1}{2} \frac{\drs(n^{i},z)}{\bar{\rho}(\bar{z})} + \frac{3}{2(1+\bar{z})} \dzs(n^{i},z) \\
&\qquad\qquad\qquad+ \frac{3}{(1+\bar{z})^{2}}\left[ \dzf(n^{i},z) \right]^{2}+\frac{\dd\delta^{(1)}\rho}{\dd\bar{z}}\frac{\dzf\left(n^{i},z\right)}{\bar{\rho}}, \notag
\end{align}
where the full expression in terms of the metric potentials is given in Appendix \ref{metriccomp}, Eq.~\eqref{eq:ddz2}. Finally, combining \eqs{eq:dvv2} and \eqref{eq:ddz2} we find the galaxy number density fluctuation at second order as defined in \eq{eq:defdg2} is 
\begin{align}
\label{eq:deltag2}
&\Delta^{(2)}_{g}(n^{i},z) =\frac{1}{2} \frac{\drs(n^{i},z)}{\bar{\rho(z)}} + \frac{3}{2}\Bigg[ -\frac{1}{2}\fs|^{s}_{o} - \frac{1}{2}\left( v_{2i} n^{i}\right)^{s}_{o}   \\
& + \frac{1}{2}\int_{0}^{\chi_{s}}\dd\chi \left( \fs' + \ps' \right)+\frac{1}{2}\left( v_{1k}v^{k}_{1}\right)^{s}_{o} - \frac{3}{2} \left( \ff|^{s}_{o}\right)^{2}+6 \ff|_{s}\ff|_{o} +\ff|_{o}\left( v_{1i}n^{i} \right)_{s} \notag \\
&-\ff|_{s}\left( v_{1i}n^{i} \right)_{o} -\left( v_{1i}n^{i} \right)_{o}\left( v_{1i}n^{i}\right)^{s}_{o}  - 2 \pf|_{o}\left( v_{1i}n^{i} \right)_{o}+ \ff|^{s}_{o}\int_{0}^{\chi_{s}}\dd\chi\left( \ff' + \pf' \right)  \notag \\
& + \left( v_{1i}n^{i} \right)_{o}\int_{0}^{\chi_{s}}\dd\chi\left( \ff' + \pf' \right)+ \ff \int_{0}^{\chi_{s}}\dd\chi\left( \ff' + \pf' \right) +  2v_{1 i}\int_{0}^{\chi_{s}}\dd \chi\left( {\ff_{,}}^{i} + {\pf_{,}}^{i} \right) \notag \\
& + 2\int_{0}^{\chi_{s}}\dd\chi\left[\pf' \left( \ff + \pf \right)\right] + 2\int_{0}^{\lambda_{s}}\dd\chi \left[ \ff \left( \frac{\dd \ff}{\dd \varsigma} - 2\frac{\dd \pf}{\dd \varsigma}\right)\right] \notag\\
& -4 \int_{0}^{\chi_{s}}\dd\chi\left[\left( \pf \ff_{,i} - \ff \pf_{,i} \right)n^{i}\right] + 4\int_{0}^{\chi_{s}}\dd\chi\left[ n^{i}\ff_{,i}\left( \ff - \pf \right)\right]  \notag \\
& + \int_{0}^{\chi_{s}}\dd\tilde{\chi}\Bigg\{2 \left( \ff' - \pf'\right)\int_{0}^{\chi_{s}}\dd\chi\left( \ff' + \pf' \right) - n^{i} \left( \ff_{,i} + \pf_{,i} \right) \int_{0}^{\chi_{s}}\dd\chi\left( \ff' + \pf'\right) \notag \\
& +  \left( \frac{\dd\ff}{\dd \varsigma} - \frac{\dd \pf}{\dd \varsigma}\right)\int_{0}^{\chi_{s}}\dd\chi \left( \ff' + \pf' \right)+2\left( \ff-\pf\right)n^{i}\int_{0}^{\chi_{s}}\dd\chi\left( \ff' + \pf'\right)_{,i} \notag \\
& +\left( \int_{0}^{\chi_{s}}\dd\chi{\left( \ff + \pf\right)_{,}}^{i}\right)\left( \int_{0}^{\chi_{s}}\dd\chi\left( \ff' + \pf'\right)_{,i}\right)\notag \\
& - n^{i}\left( \int_{0}^{\chi_{s}}\dd\chi\left( \ff' + \pf' \right)_{,i} \right)\left( \int_{0}^{\chi_{s}}\dd\chi\left( \ff' + \pf' \right) \right)\Bigg\}\Bigg]\notag \\
&+3\Bigg[ -\left( v_{1i}n^{i} + \ff\right)\big|^{s}_{o} + \int_{0}^{\chi_{s}}  \left[\ff' +\pf'\right]\dd\chi \Bigg]^{2}\notag \\
&+\frac{\dd\delta^{(1)}\rho}{\dd\bar{z}}\frac{1}{\bar{\rho}}\Bigg[ -\left( v_{1i}n^{i} + \ff\right)\big|^{s}_{o} + \int_{0}^{\chi_{s}}  \left[\ff' +\pf'\right]\dd\chi  \Bigg] \notag \\
&+\frac{1}{\H}\Bigg[\frac{1}{2}\left( \ps' -\partial_{\chi}\fs - \frac{\dd\left( v_{2i} n^{i}\right)}{\dd\varsigma}+\frac{\dd\left( v_{1k}v^{k}_{1}\right)}{\dd\varsigma}\right) \notag 
\end{align}
\begin{align}
& - 3\ff \left( \frac{\dd\ff}{\dd\varsigma}\right)+\frac{\dd\ff}{\dd\varsigma}\left( v_{1i}n^{i} \right)+\ff\frac{\dd\left( v_{1i}n^{i} \right)}{\dd\varsigma}-\left( v_{1i}n^{i} \right)\frac{\dd\left( v_{1i}n^{i}\right)}{\dd\varsigma} - 2 \frac{\dd\pf}{\dd\varsigma}\left( v_{1i}n^{i} \right) \notag\\ 
&-2\pf\frac{\dd\left( v_{1i}n^{i} \right)}{\dd\varsigma} + \frac{\dd\ff}{\dd\varsigma}\int_{0}^{\chi_{s}}\dd\chi\left( \ff' + \pf' \right) + \ff\left( \ff'+\pf'\right)\notag \\
& + \frac{\dd\left( v_{1i}n^{i} \right)}{\dd\varsigma}\int_{0}^{\chi_{s}}\dd\chi\left( \ff' + \pf' \right)+\left(v_{1i}n^{i}\right)\left( \ff'+\pf'\right) \notag \\
& + \frac{\dd\ff}{\dd\varsigma} \int_{0}^{\chi_{s}}\dd\chi\left( \ff' + \pf' \right)+\ff\left( \ff'+\pf' \right)+  2\frac{\dd v_{1 i}}{\dd\varsigma}\int_{0}^{\chi{s}}\dd \chi\left( {\ff_{,}}^{i} + {\pf_{,}}^{i} \right) + 2v_{1i}\left( {\ff_{,}}^{i} + {\pf_{,}}^{i} \right)   \notag \\
& + 2\left[\pf' \left( \ff + \pf \right)\right] + 2 \left[ \ff \left( \frac{\dd \ff}{\dd \varsigma} - 2\frac{\dd \pf}{\dd \varsigma}\right)\right]-4 \left[\left( \pf \ff_{,i} - \ff \pf_{,i} \right)n^{i}\right] \notag\\
&+ 4\left[ n^{i}\ff_{,i}\left( \ff - \pf \right)\right]+2 \left( \ff' - \pf'\right)\int_{0}^{\chi_{s}}\dd\chi\left( \ff' + \pf' \right) - n^{i} \left( \ff_{,i} + \pf_{,i} \right) \int_{0}^{\chi_{s}}\dd\chi\left( \ff' + \pf'\right) \notag \\
&  +  \left( \frac{\dd\ff}{\dd \varsigma} - \frac{\dd \pf}{\dd \varsigma}\right)\int_{0}^{\chi_{s}}\dd\chi \left( \ff' + \pf' \right)+2\left( \ff-\pf\right)n^{i}\int_{0}^{\chi_{s}}\dd\chi\left( \ff' + \pf'\right)_{,i} \notag \\
& +\left( \int_{0}^{\chi_{s}}\dd\chi{\left( \ff + \pf\right)_{,}}^{i}\right)\left( \int_{0}^{\chi_{s}}\dd\chi\left( \ff' + \pf'\right)_{,i}\right)\notag \\
& - n^{i}\left( \int_{0}^{\chi_{s}}\dd\chi\left( \ff' + \pf' \right)_{,i} \right)\left( \int_{0}^{\chi_{s}}\dd\chi\left( \ff' + \pf' \right) \right) \Bigg]\notag\\
&-\frac{1}{2}\frac{\H'}{\H^{2}}\Bigg[ -\frac{1}{2}\fs|^{s}_{o} - \frac{1}{2}\left( v_{2i} n^{i}\right)^{s}_{o} + \frac{1}{2}\int_{0}^{\chi_{s}}\dd\chi \left( \fs' + \ps' \right) +\frac{1}{2}\left( v_{1k}v^{k}_{1}\right)^{s}_{o} \notag  \\
& - \frac{3}{2} \left( \ff|^{s}_{o}\right)^{2}+6 \ff|_{s}\ff|_{o} +\ff|_{o}\left( v_{1i}n^{i} \right)_{s} -\ff|_{s}\left( v_{1i}n^{i} \right)_{o} -\left( v_{1i}n^{i} \right)_{o}\left( v_{1i}n^{i}\right)^{s}_{o}  \notag \\
& - 2 \pf|_{o}\left( v_{1i}n^{i} \right)_{o}+ \ff|^{s}_{o}\int_{0}^{\chi_{s}}\dd\chi\left( \ff' + \pf' \right) + \left( v_{1i}n^{i} \right)_{o}\int_{0}^{\chi_{s}}\dd\chi\left( \ff' + \pf' \right) \notag \\
& + \ff \int_{0}^{\chi_{s}}\dd\chi\left( \ff' + \pf' \right) +  2v_{1 i}\int_{0}^{\chi_{s}}\dd \chi\left( {\ff_{,}}^{i} + {\pf_{,}}^{i} \right) \notag \\
& + 2\int_{0}^{\chi_{s}}\dd\chi\left[\pf' \left( \ff + \pf \right)\right] + 2\int_{0}^{\lambda_{s}}\dd\chi \left[ \ff \left( \frac{\dd \ff}{\dd \varsigma} - 2\frac{\dd \pf}{\dd \varsigma}\right)\right] \notag\\
& -4 \int_{0}^{\chi_{s}}\dd\chi\left[\left( \pf \ff_{,i} - \ff \pf_{,i} \right)n^{i}\right] + 4\int_{0}^{\chi_{s}}\dd\chi\left[ n^{i}\ff_{,i}\left( \ff - \pf \right)\right]  \notag \\
& + \int_{0}^{\chi_{s}}\dd\tilde{\chi}\Bigg\{2 \left( \ff' - \pf'\right)\int_{0}^{\chi_{s}}\dd\chi\left( \ff' + \pf' \right) - n^{i} \left( \ff_{,i} + \pf_{,i} \right) \int_{0}^{\chi_{s}}\dd\chi\left( \ff' + \pf'\right) \notag \\
& +  \left( \frac{\dd\ff}{\dd \varsigma} - \frac{\dd \pf}{\dd \varsigma}\right)\int_{0}^{\chi_{s}}\dd\chi \left( \ff' + \pf' \right)+2\left( \ff-\pf\right)n^{i}\int_{0}^{\chi_{s}}\dd\chi\left( \ff' + \pf'\right)_{,i} \notag \\
& +\left( \int_{0}^{\chi_{s}}\dd\chi{\left( \ff + \pf\right)_{,}}^{i}\right)\left( \int_{0}^{\chi_{s}}\dd\chi\left( \ff' + \pf'\right)_{,i}\right)\notag \\
& - n^{i}\left( \int_{0}^{\chi_{s}}\dd\chi\left( \ff' + \pf' \right)_{,i} \right)\left( \int_{0}^{\chi_{s}}\dd\chi\left( \ff' + \pf' \right) \right)\Bigg\} \Bigg]+\frac{3}{2\H^{2}}\Bigg[ \pf' -\partial_{\chi}\ff -\frac{\dd\left( v_{1i}n^{i}\right)}{\dd\varsigma} \Bigg]^{2}\notag
\end{align}
\begin{align}
&+\frac{1}{2}\left( \frac{\H'}{\H^{2}} \right)\left( 1+\frac{\H'}{\H^{2}}\right)\Bigg[ \left(v_{1i}n^{i}+\ff+\int_{0}^{\chi_{s}}\left( \ff'+\pf'\right)\dd\chi\right) \Bigg]^{2}\notag\\
&+\frac{1}{2\H^{2}}\left(2\left[\frac{\H'}{\H}\right]^{2}+\frac{\H''}{\H} \right)\Bigg[ \left(v_{1i}n^{i}+\ff+\int_{0}^{\chi_{s}}\left( \ff'+\pf'\right)\dd\chi\right)  \Bigg]^{2} \notag \\
&+\frac{1}{2\H}\left( 1+2\frac{\H'}{\H^{2}} \right)\left( \pf' - \partial_{\chi}\ff - \frac{\dd\left( v_{1i}n^{i} \right)}{\dd\varsigma} \right)\Bigg[ v_{1i}n^{i}+\ff+\int_{0}^{\chi_{s}}\left( \ff'+\pf'\right)\dd\chi \Bigg] \notag \\
& +\Bigg[-\pf|_{s} - \pf|_{o} - \left( 1 - \frac{1}{\H \chi_{s}}\right)\ff|_{s} + \left( 2-\frac{2}{\H \chi_{s}} \right)\left( v_{1i}n^{i}\right)_{o} + \left( 1-\frac{1}{\H \chi_{s}}\right)\left( v_{1i}n^{i}\right)_{s} \notag\\
&+ \left( 1 - \frac{1}{\H\chi_{s}}\right)\int_{0}^{\chi_{s}}\left( \ff'+\pf'\right)\dd\chi -\frac{2}{\chi_{s}}\int_{0}^{\chi_{s}}\pf\dd\chi \notag \\
& -\frac{1}{2 \chi_{s}} \int_{0}^{\chi_{s}}\dd\chi \left( \chi - \chi_{s} \right)\chi \left[ \nabla^{2}\left( \ff + \pf\right) - n^{i}n^{j}\left( \ff + \pf \right)_{,ij} - \frac{2}{\chi} \frac{\dd \dnuf}{\dd \varsigma} \right]\Bigg]^{2} \notag\\
& +\Bigg[ -\ff|^{s}_{o}+\int_{0}^{\chi_{s}}\left( \ff'+\pf' \right)\dd\chi+\ff-\left( v_{1i}n^{i} \right) \Bigg]\Bigg[ -\pf|_{s} - \pf|_{o} - \left( 1 - \frac{1}{\H \chi_{s}}\right)\ff|_{s} \notag \\
& + \left( 2-\frac{2}{\H \chi_{s}} \right)\left( v_{1i}n^{i}\right)_{o}+ \left( 1-\frac{1}{\H \chi_{s}}\right)\left( v_{1i}n^{i}\right)_{s} + \left( 1 - \frac{1}{\H\chi_{s}}\right)\int_{0}^{\chi_{s}}\left( \ff'+\pf'\right)\dd\chi  \notag \\
& -\frac{2}{\chi_{s}}\int_{0}^{\chi_{s}}\pf\dd\chi -\frac{1}{2 \chi_{s}} \int_{0}^{\chi_{s}}\dd\chi \left( \chi - \chi_{s} \right)\chi \left[ \nabla^{2}\left( \ff + \pf\right) - n^{i}n^{j}\left( \ff + \pf \right)_{,ij} - \frac{2}{\chi} \frac{\dd \dnuf}{\dd \varsigma} \right] \Bigg] \notag\\
&+\frac{1}{2}\Bigg[ \dnus + \fs - \left( v_{2 i}n^{i}\right) + 2 \ff \dnuf - 2v_{1i} \dnf^{i} + \ff^{2} + \left(v_{1 k}v_{1}^{k}\right) + 4 \pf \left( v_{1i}n^{i}\right) \Bigg]\notag\\
&+\frac{\das}{\bar{d}_{A}} +\frac{\drf(n^{i},z)}{\bar{\rho}(z)} + 3\left[\left( v_{1i}n^{i} + \ff\right)\big|^{s}_{o} + \int_{0}^{\chi_{s}}\dd\chi  \left\{\ff' +\pf'\right\}\right]\times \notag \\
&\quad\Bigg[ \frac{2}{\H}\left( \pf' - \partial_{\chi}\ff + \frac{\dd \left( v_{1i}n^{i}\right)}{\dd \varsigma}\right) - 2\left( \ff + \pf \right) -3 \left( v_{1i}n^{i}\right) \notag \\
&\quad +\left( \frac{\H'}{\H^{2}} + \frac{2}{\H \chi} \right)\left( \ff - \left( v_{1i}n^{i} \right) + \int_{0}^{\chi}\dd \tilde{\chi}\left( \ff' + \pf'\right) \right) - \frac{4}{\chi}\int_{0}^{\chi}\dd\tilde{\chi}\pf \notag \\ 
&\quad  -\frac{1}{\chi}\int_{0}^{\chi}\dd \tilde{\chi}\left( \tilde{\chi}-\chi \right)\tilde{\chi} \left\{ \nabla^{2}\left( \ff + \pf \right) - n^{i}n^{j}\left( \ff + \pf\right)_{,ij} - \frac{2}{\tilde{\chi}} \frac{\dd \dnuf}{\dd \varsigma} \right\} \notag \\ 
&\quad + 3\int_{0}^{\chi}\dd \tilde{\chi}\left( \ff' + \pf' \right)\Bigg]. \notag
\end{align}
Which is the main result of this work. In the following section we make a comparison of our result with others in the literature \cite{cc1,cc2,durrer2}.

\section{Comparison with previous work}
\label{sec:comparison}

In this section we compare our linear result given in \eq{eq:Dg1} with those in the literature given in Refs.~\cite{cc1, cc2, yoo2} which also compute second order corrections and in particular with Di Dio, et al.~\cite{durrer2}. 
We do this to verify that our result is correct, since the number counts are well established to linear order with Refs.~\cite{durrer1, antony}. In a  companion paper \cite{fuentes} we perform a comparison of the leading terms of the second order expansion of the galaxy number counts.

Our result, as given in \eq{eq:ng} is
\begin{align}
\Delta^{(1)}_{g}(n^{i},z) &= \left[\frac{\drf(n^{i},z)}{\bar{\rho}(z)} + 3\ff\right] -2\left(\ff + \pf\right) + \frac{1}{\H}\left( \pf' - \partial_{\chi}\ff + \frac{\dd \left(v_{1i}n^{i}\right)}{\dd \varsigma} \right) \notag \\
& \quad +\left( \frac{\H'}{\H^{2}} +\frac{2}{\H \chi}\right)\left[ \ff - \left( v_{1i}n^{i}\right) + \int_{0}^{\chi}\dd\tilde{\chi}\left( \ff' + \pf' \right)\right] - \frac{4}{\chi}\int_{0}^{\chi}\dd \tilde{\chi}\pf\notag \\
& \quad - \frac{1}{\chi}\int_{0}^{\chi}\dd \tilde{\chi}\left( \tilde{\chi}-\chi\right)\tilde{\chi}\left[ \nabla^{2}\left( \ff + \pf\right) - n^{i}n^{j}\left( \ff + \pf \right)_{,ij} - \frac{2}{\tilde{\chi}}\frac{\dd \dnuf}{\dd \varsigma} \right]. 
\end{align}

\subsection{Di Dio, et al.}

Rewriting the result from Ref.~\cite{durrer2}, in Poisson gauge, allowing for anisotropic stress. At first order, Ref.~\cite{durrer2} have
\begin{align}
\label{eq:didio}
\Delta^{(1)}_{\text{Di Dio}}&= \delta^{(1)}_{\rho} + \left( \frac{2}{\H r}+\frac{\H'}{\H^{2}}\right)\left[ \left( v_{1i}n^{i}\right) + \psi^{I}-\psi^{A}+2\int_{\eta_{s}}^{\eta_{o}} \dd\eta' \partial_{\eta'}\psi^{I}\right] -\psi^{I} \\
&\quad +\frac{4}{r}\int_{\eta_{s}}^{\eta_{o}}\dd\eta' \psi^{I} - \frac{2}{r}\int_{\eta_{s}}^{\eta_{o}}\dd\eta'\frac{\eta'-\eta_{s}}{\eta_{o}-\eta'}\Delta_{2}\psi^{I} + \frac{1}{\H}\left[ \partial_{\eta}\psi^{I} + \partial_{r}\left( v_{1i}n^{i} \right) \right] \notag\\
&\quad -3\psi^{A} + \frac{1}{\H}\partial_{\eta}\psi^{A}, \notag
\end{align}
where ${\cal{H}} = a'(\eta)/a(\eta)$ is the Hubble parameter, the `s' denotes {\textit{source}} and the `o' denotes {\textit{observer}}, and
\be
\psi^{I} = \frac{\psi+\phi}{2}, \qquad \text{and}, \qquad \psi^{A}=\frac{\psi - \phi}{2},
\ee
which rewriting in our notation is
\begin{align}
\Delta^{(1)}_{\text{Di Dio}}&= \delta^{(1)}_{\rho} + \left( \frac{2}{\H r}+\frac{\H'}{\H^{2}}\right)\left[ \left( v_{1i}n^{i}\right) + \ff+\int_{\eta_{s}}^{\eta_{o}} \dd\eta' \left( \ff'+\pf' \right)\right] \\
&\quad+\left( 2\pf-\ff\right)+\frac{2}{r}\int_{\eta_{s}}^{\eta_{o}}\dd\eta' \left( \ff+\pf \right) - \frac{1}{r}\int_{\eta_{s}}^{\eta_{o}}\dd\eta'\frac{\eta'-\eta_{s}}{\eta_{o}-\eta'}\Delta_{2}\left( \ff+\pf\right) \notag \\
&\quad+ \frac{1}{\H}\left[ \pf' + \partial_{r}\left( v_{1i}n^{i} \right) \right] . \notag
\end{align}

Computing the difference between \eq{eq:didio} and \eq{eq:Dg1}, we have
\begin{align}
\Delta_{g}^{(1)}-\Delta_{\text{Di Dio}}^{(1)} &\approx \Bigg[\frac{\drf(n^{i},z)}{\bar{\rho}(z)} + 3\ff\Bigg] -\Bigg[\delta^{(1)}_{\rho}\Bigg]  \\
& \quad - \frac{1}{\chi}\int_{0}^{\chi}\dd \tilde{\chi}\left( \tilde{\chi}-\chi\right)\tilde{\chi}\left[ \nabla^{2}\left( \ff + \pf\right) - n^{i}n^{j}\left( \ff + \pf \right)_{,ij} \right] \notag \\
&\quad +\Bigg[ \frac{1}{r}\int_{\eta_{s}}^{\eta_{o}}\dd\eta'\frac{\eta'-\eta_{s}}{\eta_{o}-\eta'}\Delta_{2}\left( \ff+\pf\right) \Bigg] \notag
\end{align}
where the first line cancels out from the definition of the comoving density perturbation ($\delta_{\rho}^{(1)}$), and the last integral cancels out from the definition of the angular operator $\Delta_{2}$, both given in Ref.~\cite{durrer2}, and we find,
\begin{align}
\Delta_{g}^{(1)}-\Delta_{\text{Di Dio}}^{(1)} &=  0.
\end{align}

\subsection{Bertacca, et al.}

In Refs.~\cite{cc1,cc2}, their result is written in terms of ``cosmic rulers'' and it is given by
\be
\Delta_{\text{Bertacca}}^{(1)} = \delta_{g}^{(1)} +\frac{1}{2}\hat{g}_{\mu}^{\mu (1)} + b_{e} \Delta \ln a^{(1)} + \partial_{\parallel}\Delta x_{\parallel}^{(1)} + \frac{2}{\bar{\chi}}\Delta x_{\parallel}^{(1)} - 2\kappa^{(1)} + E_{\hat{0}}^{0(1)} + E_{\hat{0}}^{\parallel (1)},
\ee
which in Poisson gauge, translates into
\begin{align}
\label{eq:bertacca}
\Delta^{(1)}_{\text{Bertacca}}&= \frac{\drf(n^{i},z)}{\bar{\rho}(n^{i},z)}  - \left(  \frac{{\cal{H}}'}{{\cal{H}}^{2}}+\frac{2}{\bar{\chi} {\cal{H}}} \right)\left[ (v_{1i}n^{i} - \ff)_{o}^{s} - 2 \int_{0}^{\bar{\chi}}\ff'\dd\tilde{\chi} \right] \\
&\qquad -\ff + \frac{\ff'}{{\cal{H}}} + \frac{4}{\bar{\chi}}\int_{0}^{\bar{\chi}}\ff \dd\tilde{\chi}-\frac{1}{{\cal{H}}}\frac{\dd}{\dd \chi} (v_{1i}n^{i})- \frac{1}{{\cal{H}}}\frac{\dd\ff}{\dd \chi} \notag\\
&\qquad-2\int_{0}^{\bar{\chi}}\dd\tilde{\chi}(\bar{\chi}-\tilde{\chi})\frac{\tilde{\chi}}{\bar{\chi}}\Big[\nabla^{2}\ff + \frac{\dd^{2} \ff}{\dd \tilde{\chi}^{2}} + \ff'' - 2 \frac{\dd \ff'}{\dd \tilde{\chi}} - \frac{2}{\bar{\chi}}\left(\frac{\dd \ff}{\dd \tilde{\chi}} - \ff' \right)\Big], \notag
\end{align}
where we omitted the terms with the evolution bias $b_{e}$. We must rewrite our own result taking $\pf = \ff$ in \eq{eq:Dg1} to make the comparison, so we have that
\begin{align}
\Delta^{(1)}_{g}(n^{i},z) &= \left[\frac{\drf(n^{i},z)}{\bar{\rho}(z)} + 3\ff\right] -4\ff + \frac{1}{\H}\left( \ff' - \partial_{\chi}\ff + \frac{\dd \left(v_{1i}n^{i}\right)}{\dd \varsigma} \right) \\
& \quad +\left( \frac{\H'}{\H^{2}} +\frac{2}{\H \chi}\right)\left[ \ff - \left( v_{1i}n^{i}\right) + 2\int_{0}^{\chi}\dd\tilde{\chi} \ff' \right] - \frac{4}{\chi}\int_{0}^{\chi}\dd \tilde{\chi}\ff\notag \\
& \quad - \frac{2}{\chi}\int_{0}^{\chi}\dd \tilde{\chi}\left( \tilde{\chi}-\chi\right)\tilde{\chi}\left[ \nabla^{2}\ff - n^{i}n^{j} \ff_{,ij} - \frac{1}{\tilde{\chi}}\frac{\dd \dnuf}{\dd \varsigma} \right]. \notag
\end{align}

Computing the difference between \eq{eq:bertacca} and \eq{eq:Dg1}, we have
\begin{align}
\Delta_{g}^{(1)} - \Delta_{\text{Bertacca}}^{(1)} &\approx \frac{1}{\H}\left( \frac{\dd \left(v_{1i}n^{i}\right)}{\dd \varsigma} \right)+\frac{1}{{\cal{H}}}\frac{\dd}{\dd \chi} (v_{1i}n^{i}),
\end{align}
where both are derivatives of first order terms with respect to background quantities, and in the background $\dd \chi = \dd\varsigma$. Note that the direction in the sky, $n^{i}$, is $\left(-n^{i}\right)$ from Refs.~\cite{cc1,cc2}, so
\be
\Delta_{g}^{(1)}-\Delta_{\text{Bertacca}}^{(1)} =0.
\ee

\subsection{Yoo \& Zaldarriaga}
The galaxy overdensity in \cite{yoo2}, is given by
\be
\delta_{g}^{\text{obs}(1)} = \delta_{g}^{\text{int}(1)} +3\delta z + \delta g + 2 \frac{\delta r}{\bar{r}_{z}} - 2 \kappa + H_{z}\frac{\partial \delta r}{\partial z} + \delta u^{0} + V_{\parallel} - e_{1} \delta z_{t_{p}} - t_{1} \delta \mathcal{D}_{L},
\ee

which in Poisson gauge, allowing for anisotropic stress takes the form
\begin{align}
\label{eq:yoo}
\Delta^{(1)}_{\text{Yoo}}&= \delta_{g}^{\text{int}(1)} + 3{\cal{H}}_{o}\delta\tau_{o} -3\ff|^{z}_{o} - 3\int_{0}^{\bar{r}_{z}}\dd \bar{r} \left( \ff' + \pf'\right) + 3\ff + 3v_{1i}n^{i}+\ff + 3\pf \notag\\
&\qquad +\frac{2}{r_{z}}\Bigg[ \delta\tau_{o} - \frac{1}{{\cal{H}}_{z}}\Big( {\cal{H}}_{o}\delta\tau_{o} -\ff|^{z}_{o}-\int_{0}^{\bar{r}_{z}}\dd \bar{r}\left( \ff' + \pf'\right) + (v_{1i}n^{i})_{o}^{z}\Big)\notag \\
&\qquad +\int_{0}^{\bar{r}_{z}}\left( \ff - \pf\right)\dd \bar{r} - 2 \kappa + H_{z}\frac{\p}{\p z}\Big( \delta\tau_{o} - \frac{1}{{\cal{H}}_{z}} \Big\{ {\cal{H}}_{o}\delta\tau_{o} -\ff|^{z}_{o}-\int_{0}^{\bar{r}_{z}}\dd \bar{r}\left( \ff'+\pf'\right) \notag \\
&\qquad  +(v_{1i}n^{i})^{z}_{o}\Big\}+\int_{0}^{\bar{r}_{z}}\dd \bar{r}\left( \ff-\pf\right)  \Big)-\ff+v_{1i}n^{i}\Bigg], 
\end{align}
where we did not use the evolution bias or the running and slope of the luminosity.

The difference between \eq{eq:yoo} and \eq{eq:Dg1} is then,
\begin{align}
\Delta_{g}^{(1)}-\Delta_{\text{Yoo}}^{(1)} &\approx \left[\frac{\drf(n^{i},z)}{\bar{\rho}(z)} + 3\ff\right]-\delta_{g}^{\text{int}(1)} -3{\cal{H}}_{o}\delta\tau_{o} -\frac{2}{r_{z}} \delta\tau_{o}  \\
& \quad - \frac{1}{\chi}\int_{0}^{\chi}\dd \tilde{\chi}\left( \tilde{\chi}-\chi\right)\tilde{\chi}\left[ \nabla^{2}\left( \ff + \pf\right) - n^{i}n^{j}\left( \ff + \pf \right)_{,ij}  \right] +\frac{4}{r_{z}}  \kappa, \notag
\end{align}
where the first line is zero from the definition of $\delta_{g}^{\text{int}(1)}$, and without loss of generality we take $\delta \tau_{o} = 0$, and the integrals in the second line cancel from the definition of $\kappa$ in Ref.~\cite{yoo2}, so
\be
\Delta_{g}^{(1)}-\Delta_{\text{Yoo}}^{(1)} =  0.
\ee

\section{Conclusions \& Future work}
\label{sec:discussion}

In this paper we have provided a new and independent approach to
calculate the galaxy number overdensity. We present the galaxy number
counts in a general form depending on the affine parameter which
allows for simple plotting along the line of sight if the potentials
are known, the potentials can be calculated either using the field
equations or N-body simulations. Future surveys will provide us with
information on large and small scales and our results will help to
analyse the data and compare theoretical number counts with observed
quantities.

We present our main result in \eq{eq:deltag2}, the galaxy number
counts up to and including second order in cosmological perturbation
theory. We use scalar perturbations in longitudinal gauge allowing for
non-zero anisotropic stress. We assume a flat FLRW background universe filled
with a pressureless fluid.

As mentioned earlier, we are not the first group to
perform this calculation. We compared our result for the galaxy number
overdensities with others published in the literature, at first
order. Since other groups use different notations and approaches,
e.g.~conformal time instead of affine parameter, we adapted the
results of the other groups to our notation in order to make a clear comparison
possible.  We find that we are in agreement at linear
order with previous works. Nevertheless, the approaches taken by other groups lead to
differences in the results at second order. Given the size of the expressions involved and the complexity of rewriting the results of the other groups, we leave for a follow up paper the comparison of second order results. In Ref.~\cite{fuentes}, we tackle this issue by performing the full comparison in an Einstein-de Sitter universe.

\acknowledgments

The authors are grateful to Pedro Carrilho, Chris Clarkson, Obinna Umeh, Roy Maartens and Julian Larena for useful discussions and comments. JF acknowledges support of studentship funded by Queen Mary University of London as well as CONACYT grant No.~603085.  KAM is supported in part by the STFC under grants ST/M001202/ and ST/P000592/1. JCH acknowledges support from research grant SEP-CONACYT CB-2016-282569. The tensor algebra package \texttt{xAct} \cite{xact} and its subpackage \texttt{xPand} \cite{xpand} were employed to derive the results presented.

\appendix
\section{Connection coefficients}
\label{connection}

The connection coefficients in a FLRW spacetime, in longitudinal gauge, up to second order are
\begin{align}
\label{eq:coeff1}
\Gamma^{0}_{00} &= {\cal H} + \ff'+\frac{1}{2}\fs'-2\ff\ff', \\
\label{eq:coeff2}
\Gamma^{0}_{0i} &= \ff_{,i} + \frac{1}{2} \fs_{,i} -2\ff\ff_{,i}, \\
\label{eq:coeff3}
\Gamma^{i}_{00} &= {\ff_{,}}^{i} + \frac{1}{2} {\fs_{,}}^{i} + 2 \pf\ff_{,}^{i},  \\
\label{eq:coeff4}
\Gamma^{i}_{j0} &= \Big[{\cal H} - \pf'-\frac{1}{2}\ps' -2 \pf\pf'\Big]\delta^{i}_{j}, \\
\label{eq:coeff5}
\Gamma^{0}_{ij} &= \Bigg[ {\cal H} - 2{\cal H} \left(\ff +\pf+ \frac{1}{2} \fs+\frac{1}{2}\ps -2 \ff\pf -2\ff^{2}  \right)    \\
& \quad  -\pf' - \frac{1}{2}\ps' +2\ff\pf'  \Bigg]\delta_{ij} ,\notag \\
\label{eq:coeff6}
\Gamma^{i}_{jk} &= -{\delta^{i}_{k}}\pf_{,j} - {\delta^{i}_{j}}\pf_{,k} + {\delta_{j k}}\pf_{,}^{i}- \frac{1}{2} \left( \delta_{k}^{i}\ps_{,j} + \delta_{j}^{i}\ps_{,k} - {\delta_{jk}}\ps_{,}^{i} \right)  \\
& \quad -2 \pf \left( \delta^{i}_{k}\pf_{, j} + \delta^{i}_{j}\pf_{, k} - \delta_{j k}\pf_{,}^{i} \right),\notag 
\end{align}
including only scalar perturbations. To translate the FLRW coefficients into Minkowski spacetime we just set ${\cal H}=0$.

\section{Perturbed Ricci Tensor $R_{\mu \nu}$}
\label{riccipert}

The perturbed Ricci tensor components in a FLRW spacetime, in longitudinal gauge, up to second order are
\begin{align}
\label{eq:Ricci00}
R_{00} &= -3{\cal H}' +3\pf'' + \nabla^{2}\ff +3{\cal H} \left( \ff'+\pf' \right) +{\cal H}\Big[ \frac{3}{2}\left( \fs'+\ps' \right) + 6\left( \pf\pf' - \ff\ff' \right)\Big] \notag\\
&\quad + 2\pf\left( 3\pf''+\nabla^{2}\ff \right) +\frac{1}{2}\Big[ 6\left( \pf' \right)^{2}-6\ff' \pf'+3\ps''+\nabla^{2}\fs \Big] - \left( \ff+\pf\right)_{,i}\ff_{,}^{i}, \\
\label{eq:Ricci0i}
R_{0j} &= 2\left( \pf' +{\cal H} \ff\right)_{,j} +{\cal H}\left( \fs_{,j} -4 \ff \ff_{,j} \right) +2 \pf' \left( 2\pf-\ff \right)_{,j} +4 \pf \pf_{,j}' + \ps'_{,j}, \notag\\
&\quad \\
\label{eq:Ricci-ij}
R_{ij} &=  \left(2 {\cal H}^{2}+{\cal H}'\right) \delta_{ij} +\Big[ \nabla^{2}\pf -\pf'' - 2\left( 2{\cal H}^{2}+{\cal H}'\right)\left(\ff+\pf\right)-{\cal H}\left( \ff'+5\pf' \right) \Big] \delta_{ij}  \notag\\
&\quad + \left(\pf-\ff\right)_{,ij} +\Big\{ \ff'\pf' + \left( 2{\cal H}^{2}-{\cal H}' \right)\left[ 4\left(\ff\right)^{2}+4\ff\pf-\fs-\ps \right]  \notag\\
&\quad +{\cal H}\left[ \ff' \left( 4\ff + 2\pf\right) +10 \ff \pf' -\frac{1}{2}\left( \fs'+5\ps' \right)\right] -\frac{1}{2}\left( \ps'' - \nabla^{2}\ps\right) \notag \\
&\quad +\left( \pf '\right)^{2} + 2\ff \pf'' + 2 \pf \nabla^{2}\pf+{\left( \ff+\pf \right)_{,}}^{i}\pf_{,i}\Big\}\delta_{ij} + \left( 3\pf -\ff \right)_{,i} \pf_{,j}\notag \\
&\quad +\left( \ff-\pf\right)_{,i}\ff_{,j} + 2 \ff \ff_{,ij} + 2 \pf \pf_{,ij} +\frac{1}{2}\left( \ps - \fs \right)_{,ij} ,
\end{align}
including only scalar perturbations. To translate the FLRW components into Minkowski spacetime we just set ${\cal H}=0$.

\section{Second order in terms of the metric potentials}
\label{metriccomp}

\subsection{Geodesic Equation}
Solving \eq{eq:geodesic} at second order gives
\begin{align}
\label{eq:dnu2}
\frac{\dd \dnus}{\dd \lambda} &= -\frac{1}{2}\left[ \frac{\dd \fs}{\dd \lambda} + \frac{\dd \ps}{\dd \lambda}\right] + 2\frac{\dd \dnf^{i}}{\dd \lambda}\dnf_{i} -2 \frac{\dd \dnuf}{\dd \lambda}\dnuf - \frac{\dd\dns^{i}}{\dd \lambda}n_{i}  \\
&\qquad -4 \frac{\dd \dnuf}{\dd \lambda} \Big[ \ff+\pf \Big] - 4 \dnuf \left[ \frac{\dd \ff}{\dd \lambda} + \frac{\dd \pf}{\dd \lambda}\right] - 4 \ff \left[ \frac{\dd \ff}{\dd \lambda} + \frac{\dd \pf}{\dd \lambda}\right] \notag \\
&\qquad -4 \frac{\dd \ff}{\dd \lambda}  \Big[ \ff+\pf \Big], \notag \\
%
\label{eq:dni2}
\frac{d \dns^{i}}{d \lambda} &=  2 \frac{\dd \ps}{\dd \lambda} n^{i} - \left[ \fs_{,}^{i}+\ps_{,}^{i}\right] - 4 \ff\pf' n^{i} + 4\dnf^{i}\pf' + 4 \ff \ff_{,}^{i}  \\
&\qquad - 2 \dnf^{j}\dnf_{,j}^{i} -2\dnf^{i} \dnuf + 4 \Big[ n^{i}\dnf^{j} + n^{j}\dnf^{i} +2 n^{j}n^{i} \pf \Big] \pf_{,j} \notag \\
&\qquad -4\pf \Big[ \pf_{,}^{i} - 2n^{i}\pf' \Big] - 4 n^{j}\dnf_{j}\Big[ \left( \ff_{,}^{i} + \pf_{,}^{i} \right) - n^{i} \pf' \Big], \notag 
\end{align}
Using \eq{eq:null-conditions}, and the integrated version of \eqs{eq:dnu1} and \eqref{eq:dn1}, we rewrite \eqs{eq:dnu2} and \eqref{eq:dni2} purely in terms of the metric potentials, 
\begin{align}
\label{eq:dnu2-metric}
\frac{\dd \dnus}{\dd \lambda} &= -2 \frac{\dd \fs}{\dd\lambda}+\fs'+\ps' + \frac{1}{2}\left[ \frac{\dd\fs}{\dd\lambda} + \frac{\dd\ps}{\dd\lambda}\right] + 4 \ff \left[3\frac{\dd\ff}{\dd \lambda} - \frac{\dd\pf}{\dd\lambda}\right] \\
&\qquad -4\ff'\left( \ff+\pf\right) + 4\ff \left( 3\ff + \pf \right) + 4 \pf \int_{\lambda_{o}}^{\lambda_{s}}\left( \ff'' + \pf'' \right)\dd \lambda \notag \\
&\qquad -\left( 4\ff + \pf \right)_{,i} \int_{\lambda_{o}}^{\lambda_{s}}{\left( \ff + \pf \right)_{,}}^{i}\dd \lambda + 4 \frac{\dd\pf}{\dd\lambda} \int_{\lambda_{o}}^{\lambda_{s}}\left( \ff' + \pf' \right)\dd \lambda \notag \\
&\qquad -2 \left[ 3\ff + \ff + \ff' + 3 \pf'\right]\int_{\lambda_{o}}^{\lambda_{s}}\left( \ff' + \pf' \right)\dd \lambda\notag \\
&\qquad + 6 \left( \int_{\lambda_{o}}^{\lambda_{s}}\left( \ff + \pf \right)_{,i} \dd \lambda \right) \left( \int_{\lambda_{o}}^{\lambda_{s}} {\left( \ff + \pf \right)_{,}}^{i} \dd \lambda \right) \notag \\
&\qquad +2 \left( \int_{\lambda_{o}}^{\lambda_{s}}\left( \ff' + \pf' \right)\dd \lambda \right) \left( \int_{\lambda_{o}}^{\lambda_{s}}\left( \ff' + \pf' \right)\dd \lambda \right), \notag 
\end{align}
\begin{align}
\label{eq:dni2-metric}
\frac{d \dns^{i}}{d \lambda} &=  2 \frac{\dd \ps}{\dd \lambda} n^{i} - \left[ \fs_{,}^{i}+\ps_{,}^{i}\right] +  24 \pf \pf' n^{i} + 8 \ff \ff_{,}^{i} -8 n^{i} \ff \pf'  \\
&\qquad   -4 \pf'  \int_{\lambda_{o}}^{\lambda_{s}}{\left( \ff + \pf\right)_{,}}^{i} +4\ff\pf_{,}^{i}-4\pf{\ff_{,}}^{i}-8\pf{\pf_{,}}^{i}+4 n^{i} \ff\ff'  \notag \\
&\qquad - 8\pf \frac{\dd \pf}{\dd\lambda}n^{i} +4\pf {\left( \ff+\pf\right)_{,}}^{i} -4 \pf \int_{\lambda_{o}}^{\lambda_{s}}{\left( \ff + \pf\right)_{,}}^{i}d\lambda \notag \\
&\qquad + 4 n^{i} \pf_{,j}\int_{\lambda_{o}}^{\lambda_{s}}{\left( \ff + \pf\right)_{,}}^{j}d\lambda -2 \left( \int_{\lambda_{o}}^{\lambda_{s}}{\left( \ff + \pf\right)_{,}}^{j}d\lambda \right)\left( \int_{\lambda_{o}}^{\lambda_{s}}{{\left( \ff + \pf\right)_{,}}^{i}}_{j}d\lambda \right) \notag \\
&\qquad + 8 \ff\pf n^{i} -4 \pf n^{i}\int_{\lambda_{o}}^{\lambda_{s}}\left( \ff' + \pf' \right)\dd\lambda -4\ff \int_{\lambda_{o}}^{\lambda_{s}}{\left( \ff+\pf \right)_{,}}^{i} \dd\lambda \notag \\
&\qquad +2 \left(\int_{\lambda_{o}}^{\lambda_{s}}\left( \ff' + \pf' \right)\dd\lambda\right)\left(\int_{\lambda_{o}}^{\lambda_{s}}{\left( \ff+\pf \right)_{,}}^{i} \dd\lambda \right)  \notag \\
&\qquad + 12 n^{i} \pf \left[ \frac{\dd\pf}{\dd\lambda}-\pf'\right] - 4n^{i}\pf_{,j} \int_{\lambda_{o}}^{\lambda_{s}}{\left( \ff+\pf \right)_{,}}^{j} d\lambda \notag \\
&\qquad  - \left[ \frac{\dd\pf}{\dd\lambda}-\pf'\right] \int_{\lambda_{o}}^{\lambda_{s}}{\left( \ff+\pf \right)_{,}}^{i} \dd\lambda -4 \Big[ \left( \ff_{,}^{i} + \pf_{,}^{i} \right) - n^{i} \pf' \Big]\int_{\lambda_{o}}^{\lambda_{s}}\left( \ff' + \pf' \right)\dd\lambda, \notag 
\end{align}
where we integrate along the line of sight from $\lambda_{o}$ to $\lambda_{s}$.

\subsection{Energy}
The perturbed energy in terms of the metric potentials is given by 
\begin{align}
\label{eq:dE1-metric}
\Ef &= \left(-2 \ff \Big|_{o}^{s}+\int_{\lambda_{o}}^{\lambda^{s}}\left( \ff' +\pf' \right)\dd\lambda\right) + \ff - v_{1i}n^{i},  \\
\label{eq:dE2-metric}
\Es &=  \frac{1}{2}\Bigg[ 2 \Big\{  \left(-2 \ff \Big|_{o}^{s}+\int_{\lambda_{o}}^{\lambda^{s}}\left( \ff' +\pf' \right)\dd\lambda\right) + \ff - v_{1i}n^{i}  \Big\} \ff  \\
&\qquad - \ff^{2}-5\pf^{2}- 6\ff\pf + \frac{1}{2}\left( \fs-\ps\right) - n^{i}v_{2i} \notag \\
&\qquad - 4\left( \ff + \pf\right) \ff\Big|^{s}_{o}+ 2\left( \ff + \pf\right)\int_{\lambda_{o}}^{\lambda_{s}}\left( \ff' + \pf' \right)\dd\lambda  \notag \\
&\qquad  -4n^{i}v_{1 i} \pf\Big|^{s}_{o} +2 v_{1 i}  \int_{\lambda_{o}}^{\lambda_{s}}{\left( \ff + \pf \right)_{,}}^{i} \dd\lambda \notag \\
&\qquad +\int_{\lambda_{o}}^{\lambda_{s}}\Bigg\{-2 \frac{\dd \fs}{\dd\lambda}+\fs'+\ps' + \frac{1}{2}\left[ \frac{\dd\fs}{\dd\lambda} + \frac{\dd\ps}{\dd\lambda}\right] + 4 \ff \left[3\frac{\dd\ff}{\dd \lambda} - \frac{\dd\pf}{\dd\lambda}\right] \notag \\
&\qquad -4\ff'\left( \ff+\pf\right) + 4\ff \left( 3\ff + \pf \right) + 4 \pf \int_{\lambda_{o}}^{\lambda_{s}}\left( \ff'' + \pf'' \right)\dd \lambda \notag \\
&\qquad -\left( 4\ff + \pf \right)_{,i} \int_{\lambda_{o}}^{\lambda_{s}}{\left( \ff + \pf \right)_{,}}^{i}\dd \lambda + 4 \frac{\dd\pf}{\dd\lambda} \int_{\lambda_{o}}^{\lambda_{s}}\left( \ff' + \pf' \right)\dd \lambda \notag \\
&\qquad -2 \left[ 3\ff + \ff + \ff' + 3 \pf'\right]\int_{\lambda_{o}}^{\lambda_{s}}\left( \ff' + \pf' \right)\dd \lambda\notag \\
&\qquad + 6 \left( \int_{\lambda_{o}}^{\lambda_{s}}\left( \ff + \pf \right)_{,i} \dd \lambda \right) \left( \int_{\lambda_{o}}^{\lambda_{s}} {\left( \ff + \pf \right)_{,}}^{i} \dd \lambda \right) \notag \\
&\qquad +2 \left( \int_{\lambda_{o}}^{\lambda_{s}}\left( \ff' + \pf' \right)\dd \lambda \right) \left( \int_{\lambda_{o}}^{\lambda_{s}}\left( \ff' + \pf' \right)\dd \lambda \right)\Bigg\} \dd \tilde{\lambda} \Bigg]. \notag
\end{align}

\subsection{Observed Redshift}
At second order the redshift is
\begin{align}
\label{eq:dz2}
\dzs&= \left[ \ff^{2} + \left( v_{1i}n^{i}\right)^{2} -2 \ff\left( v_{1i}n^{i}\right) \right]_{o}-\ff|_{o} \dnuf|_{s} + \dnuf|_{s} \left( v_{1i}n^{i}\right)_{o} -\ff|_{s} \ff|_{o} \\
&\qquad + \ff|_{s}\left( v_{1i}n^{i}\right)_{o}+\left( v_{1i}n^{i}\right)_{s}\ff|_{o}-\left( v_{1i}n^{i}\right)_{s}\left( v_{1i}n^{i}\right)_{o} + \frac{1}{2}\Bigg[ 2\Ef \ff  - \ff^{2}-5\pf^{2} \notag\\
&\qquad - 6\ff\pf +\dnus + \frac{1}{2}\left( \fs-\ps\right) +2 \dnuf \left( \ff + \pf\right)  - 2\dnf^{i}v_{1 i} - n^{i}v_{2i} \Bigg]_{s} \notag\\
&\qquad - \frac{1}{2}\Bigg[ \ff^{2}-5\pf^{2}- 6\ff\pf + \frac{1}{2}\left( \fs-\ps\right) - (v_{1 i}n^{i}) \ff - n^{i}v_{2i} \Bigg]_{o}, \notag
\end{align}
and using \eqs{eq:dnu1}, \eqref{eq:dn1}, \eqref{eq:dnu2} and \eqref{eq:dE1-metric}, in terms of the metric potentials is
\begin{align}
\label{eq:dz2-metric}
\dzs&= \left[ \ff^{2} + \left( v_{1i}n^{i}\right)^{2} -2 \ff\left( v_{1i}n^{i}\right) \right]_{o}-\ff|_{o} \left[ -2\ff + \int_{\lambda_{o}}^{\lambda_{s}}\left( \ff' + \pf' \right)\dd\lambda \right]_{s} \\
&\qquad + \left[ -2\ff + \int_{\lambda_{o}}^{\lambda_{s}}\left( \ff' + \pf' \right)\dd\lambda \right]_{s} \left( v_{1i}n^{i}\right)_{o} -\ff|_{s} \ff|_{o} \notag \\
&\qquad + \ff|_{s}\left( v_{1i}n^{i}\right)_{o}+\left( v_{1i}n^{i}\right)_{s}\ff|_{o}-\left( v_{1i}n^{i}\right)_{s}\left( v_{1i}n^{i}\right)_{o} \notag \\
&\qquad + \frac{1}{2}\Bigg[ 2 \Big\{  \left(-2 \ff \Big|_{o}^{s}+\int_{\lambda_{o}}^{\lambda^{s}}\left( \ff' +\pf' \right)\dd\lambda\right) + \ff - v_{1i}n^{i}  \Big\} \ff \notag \\
&\qquad - \ff^{2}-5\pf^{2}- 6\ff\pf + \frac{1}{2}\left( \fs-\ps\right) - n^{i}v_{2i} \notag \\
&\qquad - 4\left( \ff + \pf\right) \ff\Big|^{s}_{o}+ 2\left( \ff + \pf\right)\int_{\lambda_{o}}^{\lambda_{s}}\left( \ff' + \pf' \right)\dd\lambda  \notag \\
&\qquad  -4n^{i}v_{1 i} \pf\Big|^{s}_{o} +2 v_{1 i}  \int_{\lambda_{o}}^{\lambda_{s}}{\left( \ff + \pf \right)_{,}}^{i} \dd\lambda \notag \\
&\qquad +\int_{\lambda_{o}}^{\lambda_{s}}\Bigg\{-2 \frac{\dd \fs}{\dd\lambda}+\fs'+\ps' + \frac{1}{2}\left[ \frac{\dd\fs}{\dd\lambda} + \frac{\dd\ps}{\dd\lambda}\right] + 4 \ff \left[3\frac{\dd\ff}{\dd \lambda} - \frac{\dd\pf}{\dd\lambda}\right] \notag \\
&\qquad -4\ff'\left( \ff+\pf\right) + 4\ff \left( 3\ff + \pf \right) + 4 \pf \int_{\lambda_{o}}^{\lambda_{s}}\left( \ff'' + \pf'' \right)\dd \lambda \notag \\
&\qquad -\left( 4\ff + \pf \right)_{,i} \int_{\lambda_{o}}^{\lambda_{s}}{\left( \ff + \pf \right)_{,}}^{i}\dd \lambda + 4 \frac{\dd\pf}{\dd\lambda} \int_{\lambda_{o}}^{\lambda_{s}}\left( \ff' + \pf' \right)\dd \lambda \notag \\
&\qquad -2 \left[ 3\ff + \ff + \ff' + 3 \pf'\right]\int_{\lambda_{o}}^{\lambda_{s}}\left( \ff' + \pf' \right)\dd \lambda\notag \\
&\qquad + 6 \left( \int_{\lambda_{o}}^{\lambda_{s}}\left( \ff + \pf \right)_{,i} \dd \lambda \right) \left( \int_{\lambda_{o}}^{\lambda_{s}} {\left( \ff + \pf \right)_{,}}^{i} \dd \lambda \right) \notag \\
&\qquad +2 \left( \int_{\lambda_{o}}^{\lambda_{s}}\left( \ff' + \pf' \right)\dd \lambda \right) \left( \int_{\lambda_{o}}^{\lambda_{s}}\left( \ff' + \pf' \right)\dd \lambda \right)\Bigg\} \dd \tilde{\lambda} \Bigg]_{s} \notag\\
&\qquad - \frac{1}{2}\Bigg[ \ff^{2}-5\pf^{2}- 6\ff\pf + \frac{1}{2}\left( \fs-\ps\right) - (v_{1 i}n^{i}) \ff - n^{i}v_{2i}  \Bigg]_{o}. \notag
\end{align}

\subsection{Angular Diameter Distance}
Using \eqs{eq:dnu1}, \eqref{eq:dn1}, \eqref{eq:dnu2-metric}, \eqref{eq:dni2-metric}, \eqref{eq:dE2-metric}, \eqref{eq:da-linear}, \eqref{eq:Ricci00}, \eqref{eq:Ricci0i}, \eqref{eq:Ricci-ij} and \eqref{eq:contracted-null-shear} we find that the second order perturbation to the angular diameter distance becomes
\begin{align}
\label{eq:da-second-metric}
\frac{\das(\lambda_{s}) }{\bar{d}_{A}(\lambda_{s}) }&= \frac{1}{2}\Big[\ff^{2}-5\pf^{2}- 6\ff\pf + \frac{1}{2}\left( \fs-\ps\right) - (v_{1 i}n^{i}) \ff - n^{i}v_{2i} \Big]_{o} \\
&\qquad -\frac{1}{\lambda_{o}-\lambda_{s}}\int_{\lambda_{o}}^{\lambda_{s}}\dd\lambda\int_{\lambda_{s}}^{\lambda_{s}}\dd\tilde{\lambda}\Big\{4\left( -2\ff\Big|^{s}_{s} + \int_{\lambda_{o}}^{\lambda_{s}}\left( \ff'+\pf' \right)\dd \lambda \right)  \Bigg(  \ff |_{o} - (v_{1 i}n^{i})_{o}-\ff|_{o}^{s}\notag \\
&\qquad -\frac{3}{2}\left( \pf-\ff\right)\Big|_{o}^{s}+\int_{\lambda_{o}}^{\lambda_{s}}\dd\lambda\int_{\lambda_{s}}^{\lambda_{s}}\dd\tilde{\lambda}\left\{ \left( \ff+\pf\right)'' - \nabla^{2}\left( \ff+\pf \right) \right\} \notag \\
&\qquad -\frac{1}{\bar{d}_{A}}\left[ 2\int_{\lambda_{o}}^{\lambda_{s}}\dd\lambda\int_{\lambda_{s}}^{\lambda_{s}}\dd\tilde{\lambda}\left( \ff+\pf\right)'-\int_{\lambda_{o}}^{\lambda_{s}}\dd\lambda\int_{\lambda_{s}}^{\lambda_{s}}\dd\tilde{\lambda}\int_{\lambda_{o}}^{\lambda_{s}}\dd\breve{\lambda}\left\{ \left( \ff+\pf \right)''-\nabla^{2}\left( \ff+\pf \right) \right\} \right] \Bigg)'' \notag \\
&\qquad +\Bigg( \ff |_{o} - (v_{1 i}n^{i})_{o}-\ff|_{o}^{s}-\frac{3}{2}\left( \pf-\ff\right)\Big|_{o}^{s}+\int_{\lambda_{o}}^{\lambda_{s}}\dd\lambda\int_{\lambda_{s}}^{\lambda_{s}}\dd\tilde{\lambda}\left\{ \left( \ff+\pf\right)'' - \nabla^{2}\left( \ff+\pf \right) \right\} \notag \\
&\qquad -\frac{1}{\bar{d}_{A}}\left[ 2\int_{\lambda_{o}}^{\lambda_{s}}\dd\lambda\int_{\lambda_{s}}^{\lambda_{s}}\dd\tilde{\lambda}\left( \ff+\pf\right)'-\int_{\lambda_{o}}^{\lambda_{s}}\dd\lambda\int_{\lambda_{s}}^{\lambda_{s}}\dd\tilde{\lambda}\int_{\lambda_{o}}^{\lambda_{s}}\dd\breve{\lambda}\left\{ \left( \ff+\pf \right)''-\nabla^{2}\left( \ff+\pf \right) \right\} \right]\Bigg)' \times \notag \\
&\qquad \times \left( - 2 \frac{\dd \ff}{\dd \lambda} + \ff' + \pf' \right) +\frac{1}{2} \Bigg( \ff |_{o} - (v_{1 i}n^{i})_{o}-\ff|_{o}^{s}-\frac{3}{2}\left( \pf-\ff\right)\Big|_{o}^{s}\notag \\
&\qquad +\int_{\lambda_{o}}^{\lambda_{s}}\dd\lambda\int_{\lambda_{s}}^{\lambda_{s}}\dd\tilde{\lambda}\left\{ \left( \ff+\pf\right)'' - \nabla^{2}\left( \ff+\pf \right) \right\} \notag \\
&\qquad -\frac{1}{\lambda_{o}-\lambda_{s}}\left[ 2\int_{\lambda_{o}}^{\lambda_{s}}\dd\lambda\int_{\lambda_{s}}^{\lambda_{s}}\dd\tilde{\lambda}\left( \ff+\pf\right)'-\int_{\lambda_{o}}^{\lambda_{s}}\dd\lambda\int_{\lambda_{s}}^{\lambda_{s}}\dd\tilde{\lambda}\int_{\lambda_{o}}^{\lambda_{s}}\dd\breve{\lambda}\left\{ \left( \ff+\pf \right)''-\nabla^{2}\left( \ff+\pf \right) \right\} \right]\Bigg) \times \notag \\
&\qquad \Bigg( \left[ \frac{\dd^{2}\pf}{\dd\lambda^{2}}- \frac{\dd^{2}\ff}{\dd\lambda^{2}} \right]  +2\left[ \frac{\dd\ff'}{\dd\lambda}+ \frac{\dd\pf'}{\dd\lambda}  \right] + \nabla^{2}\left( \ff+\pf\right) - \left(\ff''+\pf''\right) \Bigg) \Big\}\notag \\
&\qquad +\frac{1}{\lambda_{o}-\lambda_{s}}\int_{\lambda_{o}}^{\lambda_{s}}\dd\lambda\int_{\lambda_{s}}^{\lambda_{s}}\dd\tilde{\lambda}\Bigg\{ \frac{1}{2} \Bigg[ -2 \frac{\dd \fs}{\dd\lambda}+\fs'+\ps' + \frac{1}{2}\left[ \frac{\dd\fs}{\dd\lambda} + \frac{\dd\ps}{\dd\lambda}\right] + 4 \ff \left[3\frac{\dd\ff}{\dd \lambda} - \frac{\dd\pf}{\dd\lambda}\right] \notag \\
&\qquad -4\ff'\left( \ff+\pf\right) + 4\ff \left( 3\ff + \pf \right) + 4 \pf \int_{\lambda_{o}}^{\lambda_{s}}\left( \ff'' + \pf'' \right)\dd \lambda \notag \\
&\qquad -\left( 4\ff + \pf \right)_{,i} \int_{\lambda_{o}}^{\lambda_{s}}{\left( \ff + \pf \right)_{,}}^{i}\dd \lambda + 4 \frac{\dd\pf}{\dd\lambda} \int_{\lambda_{o}}^{\lambda_{s}}\left( \ff' + \pf' \right)\dd \lambda \notag \\
&\qquad -2 \left[ 3\ff + \ff + \ff' + 3 \pf'\right]\int_{\lambda_{o}}^{\lambda_{s}}\left( \ff' + \pf' \right)\dd \lambda\notag \\
&\qquad + 6 \left( \int_{\lambda_{o}}^{\lambda_{s}}\left( \ff + \pf \right)_{,i} \dd \lambda \right) \left( \int_{\lambda_{o}}^{\lambda_{s}} {\left( \ff + \pf \right)_{,}}^{i} \dd \lambda \right) \notag \\
&\qquad +2 \left( \int_{\lambda_{o}}^{\lambda_{s}}\left( \ff' + \pf' \right)\dd \lambda \right) \left( \int_{\lambda_{o}}^{\lambda_{s}}\left( \ff' + \pf' \right)\dd \lambda \right) \Bigg] \notag \\
&\qquad - \left( -2\ff\Big|^{s}_{s} + \int_{\lambda_{o}}^{\lambda_{s}}\left( \ff'+\pf' \right)\dd \lambda \right)\left( -2\ff\Big|^{s}_{s} + \int_{\lambda_{o}}^{\lambda_{s}}\left( \ff'+\pf' \right)\dd \lambda \right)' \notag \\
&\qquad +\left( -2n^{i}\pf\Big|^{s}_{s} + \int_{\lambda_{o}}^{\lambda_{s}}{\left( \ff+\pf \right)_{,}}^{i}\dd \lambda \right)\left( -2\ff\Big|^{s}_{s} + \int_{\lambda_{o}}^{\lambda_{s}}\left( \ff'+\pf' \right)\dd \lambda \right)_{,i}\Bigg\}\notag 
\end{align}
\begin{align}
&\qquad -\frac{1}{2}\int_{\lambda_{o}}^{\lambda_{s}}\dd\lambda\int_{\lambda_{s}}^{\lambda_{s}}\dd\tilde{\lambda}\Bigg\{ \frac{1}{2}\left[ \frac{\dd^{2}\pf}{\dd\lambda^{2}}-\frac{\dd^{2}\ff}{\dd\lambda^{2}} \right] + \left[ \frac{\dd\ff}{\dd\lambda}+\frac{\dd\pf}{\dd\lambda}\right] - \frac{1}{2}\left( \ff'' + \pf'' \right) + \frac{1}{2}\nabla^{2} \left( \fs + \ps \right) \notag \\
&\qquad +\left(\ff'\right)^{2} - \left( \pf '\right)^{2} + 2 \ff \left[ \ff''-\nabla^{2}\ff - \frac{\dd^{2}\pf}{\dd\lambda^{2}} - 3 \frac{\dd\ff'}{\dd\lambda} \right] \notag \\
&\qquad - 2 \pf \left[ \pf''-\nabla^{2}\pf - \frac{\dd^{2}\ff}{\dd\lambda^{2}} - 3 \frac{\dd\pf'}{\dd\lambda} \right] -\ff_{,i}{\ff_{,}}^{i}+\pf_{,i}{\pf_{,}}^{i}\notag \\
&\qquad +4\pf\left[ \frac{\dd^{2}\pf}{\dd\lambda^{2}} - \frac{\dd^{2}\ff}{\dd\lambda^{2}} +\frac{\dd \ff'}{\dd\lambda} - \frac{\dd \pf'}{\dd\lambda} + \frac{\dd \ff}{\dd\lambda}+ \frac{\dd \pf}{\dd\lambda} -\ff'-\pf' \right] \notag\\
&\qquad -2 \left[ \frac{\dd \pf}{\dd\lambda} - \frac{\dd \ff}{\dd\lambda} +\ff + \pf \right]_{,i}\int_{\lambda_{o}}^{\lambda_{s}}{\left( \ff + \pf\right)_{,}}^{i} + 4\ff \frac{\dd\pf}{\dd\lambda} + 4\pf \frac{\dd\pf}{\dd\lambda} \notag \\
&\qquad +2 \left( \ff+\pf\right)\nabla^{2}\left( \ff+\pf \right) + 4\frac{\dd\pf}{\dd\lambda}\left[ -2\ff\Big|^{s}_{o} + \int_{\lambda_{o}}^{\lambda_{s}}\left( \ff' + \pf'\right)\dd\lambda\right] \notag \\
&\qquad + 2 \nabla^{2}\left( \ff+\pf\right)\left[ -2\ff\Big|^{s}_{o} + \int_{\lambda_{o}}^{\lambda_{s}}\left( \ff' + \pf'\right)\dd\lambda\right]\notag \\
&\qquad -\frac{3}{2} \Bigg[ \frac{\dd}{\dd \lambda}\left( \ff + \pf \right) -2 \left( \ff' + \pf' \right) + \int_{\lambda_{o}}^{\lambda_{s}}\left( \ff''+\pf'' \right)\dd\lambda \Bigg]^{2} \notag \\
& \qquad + 8 \left( \ff' + \pf' \right)^{2} + 2 \left[ \int_{\lambda_{o}}^{\lambda_{s}}\left( \ff'' + \pf'' \right)\dd\lambda\right]^{2} + 2 \left[ \frac{\dd \ff}{\dd\lambda} + \frac{\dd \pf}{\dd \lambda} \right]^{2}\notag \\
&\qquad - \left( \ff+\pf \right) \Bigg[ 8\frac{\dd}{\dd\lambda}\left( \ff'' + \pf'' \right) + 4\left( \ff'' + \pf'' \right) + \nabla^{2}\left( \ff + \pf \right)  \Bigg] \notag \\
&\qquad - \left( \ff' + \pf' \right) \int_{\lambda_{o}}^{\lambda_{s}}\Big[ 8\left( \ff'' + \pf'' \right)-2\nabla^{2}\left( \ff + \pf \right) \Big]\dd\lambda \notag \\
&\qquad +\left(\int_{\lambda_{o}}^{\lambda_{s}}\left( \ff+\pf \right)_{,ij}\dd\lambda \right) \left(\int_{\lambda_{o}}^{\lambda_{s}}{\left( \ff+\pf \right)_{,}}^{ij}\dd\lambda\right) \notag \\
&\qquad +\left(\int_{\lambda_{o}}^{\lambda_{s}}\nabla^{2}\left( \ff+\pf \right)\dd\lambda \right) \left(\int_{\lambda_{o}}^{\lambda_{s}}\left( \ff'' + \pf'' \right)\dd\lambda\right) \notag \\
&\qquad -\left(\int_{\lambda_{o}}^{\lambda_{s}}\nabla^{2}\left( \ff+\pf \right)\dd\lambda \right) \left(\int_{\lambda_{o}}^{\lambda_{s}}\nabla^{2}\left( \ff + \pf \right)\dd\lambda\right) \notag \\
&\qquad - 2\left( \ff+\pf \right)_{,i}{\left( \ff+\pf \right)_{,}}^{i}  + 4 {\left( \ff+\pf \right)_{,}}^{i} \Bigg[ \int_{\lambda_{o}}^{\lambda_{s}}\left( \ff+\pf \right)_{,i} \dd\lambda\Bigg]\notag \\
&\qquad - 2 \left( \int_{\lambda_{o}}^{\lambda_{s}}\left( \ff+\pf \right)_{,i} \dd\lambda \right)\left( \int_{\lambda_{o}}^{\lambda_{s}}{\left( \ff+\pf \right)_{,}}^{i} \dd\lambda \right)\Bigg\}\notag 
\end{align}
\begin{align}
&\qquad+\frac{2}{\lambda_{o}-\lambda_{s}} \int_{\lambda_{o}}^{\lambda_{s}}\dd\lambda\int_{\lambda_{s}}^{\lambda_{s}}\dd\tilde{\lambda}\int_{\lambda_{o}}^{\lambda_{s}}\dd\breve{\lambda}\Bigg\{ \frac{1}{2}\left[ \frac{\dd^{2}\pf}{\dd\lambda^{2}}-\frac{\dd^{2}\ff}{\dd\lambda^{2}} \right] + \left[ \frac{\dd\ff}{\dd\lambda}+\frac{\dd\pf}{\dd\lambda}\right] - \frac{1}{2}\left( \ff'' + \pf'' \right) + \frac{1}{2}\nabla^{2} \left( \fs + \ps \right) \notag \\
&\qquad +\left(\ff'\right)^{2} - \left( \pf '\right)^{2} + 2 \ff \left[ \ff''-\nabla^{2}\ff - \frac{\dd^{2}\pf}{\dd\lambda^{2}} - 3 \frac{\dd\ff'}{\dd\lambda} \right] \notag \\
&\qquad - 2 \pf \left[ \pf''-\nabla^{2}\pf - \frac{\dd^{2}\ff}{\dd\lambda^{2}} - 3 \frac{\dd\pf'}{\dd\lambda} \right] -\ff_{,i}{\ff_{,}}^{i}+\pf_{,i}{\pf_{,}}^{i}\notag \\
&\qquad +4\pf\left[ \frac{\dd^{2}\pf}{\dd\lambda^{2}} - \frac{\dd^{2}\ff}{\dd\lambda^{2}} +\frac{\dd \ff'}{\dd\lambda} - \frac{\dd \pf'}{\dd\lambda} + \frac{\dd \ff}{\dd\lambda}+ \frac{\dd \pf}{\dd\lambda} -\ff'-\pf' \right] \notag\\
&\qquad -2 \left[ \frac{\dd \pf}{\dd\lambda} - \frac{\dd \ff}{\dd\lambda} +\ff + \pf \right]_{,i}\int_{\lambda_{o}}^{\lambda_{s}}{\left( \ff + \pf\right)_{,}}^{i} + 4\ff \frac{\dd\pf}{\dd\lambda} + 4\pf \frac{\dd\pf}{\dd\lambda} \notag \\
&\qquad +2 \left( \ff+\pf\right)\nabla^{2}\left( \ff+\pf \right) + 4\frac{\dd\pf}{\dd\lambda}\left[ -2\ff\Big|^{s}_{o} + \int_{\lambda_{o}}^{\lambda_{s}}\left( \ff' + \pf'\right)\dd\lambda\right] \notag \\
&\qquad + 2 \nabla^{2}\left( \ff+\pf\right)\left[ -2\ff\Big|^{s}_{o} + \int_{\lambda_{o}}^{\lambda_{s}}\left( \ff' + \pf'\right)\dd\lambda\right] \notag \\
&\qquad -\frac{3}{2} \Bigg[ \frac{\dd}{\dd \lambda}\left( \ff + \pf \right) -2 \left( \ff' + \pf' \right) + \int_{\lambda_{o}}^{\lambda_{s}}\left( \ff''+\pf'' \right)\dd\lambda \Bigg]^{2} \notag \\
& \qquad + 8 \left( \ff' + \pf' \right)^{2} + 2 \left[ \int_{\lambda_{o}}^{\lambda_{s}}\left( \ff'' + \pf'' \right)\dd\lambda\right]^{2} + 2 \left[ \frac{\dd \ff}{\dd\lambda} + \frac{\dd \pf}{\dd \lambda} \right]^{2}\notag \\
&\qquad - \left( \ff+\pf \right) \Bigg[ 8\frac{\dd}{\dd\lambda}\left( \ff'' + \pf'' \right) + 4\left( \ff'' + \pf'' \right) + \nabla^{2}\left( \ff + \pf \right)  \Bigg] \notag \\
&\qquad - \left( \ff' + \pf' \right) \int_{\lambda_{o}}^{\lambda_{s}}\Big[ 8\left( \ff'' + \pf'' \right)-2\nabla^{2}\left( \ff + \pf \right) \Big]\dd\lambda \notag \\
&\qquad +\left(\int_{\lambda_{o}}^{\lambda_{s}}\left( \ff+\pf \right)_{,ij}\dd\lambda \right) \left(\int_{\lambda_{o}}^{\lambda_{s}}{\left( \ff+\pf \right)_{,}}^{ij}\dd\lambda\right) \notag \\
&\qquad +\left(\int_{\lambda_{o}}^{\lambda_{s}}\nabla^{2}\left( \ff+\pf \right)\dd\lambda \right) \left(\int_{\lambda_{o}}^{\lambda_{s}}\left( \ff'' + \pf'' \right)\dd\lambda\right) \notag \\
&\qquad -\left(\int_{\lambda_{o}}^{\lambda_{s}}\nabla^{2}\left( \ff+\pf \right)\dd\lambda \right) \left(\int_{\lambda_{o}}^{\lambda_{s}}\nabla^{2}\left( \ff + \pf \right)\dd\lambda\right) \notag \\
&\qquad - 2\left( \ff+\pf \right)_{,i}{\left( \ff+\pf \right)_{,}}^{i}  + 4 {\left( \ff+\pf \right)_{,}}^{i} \Bigg[ \int_{\lambda_{o}}^{\lambda_{s}}\left( \ff+\pf \right)_{,i} \dd\lambda\Bigg]\notag \\
&\qquad - 2 \left( \int_{\lambda_{o}}^{\lambda_{s}}\left( \ff+\pf \right)_{,i} \dd\lambda \right)\left( \int_{\lambda_{o}}^{\lambda_{s}}{\left( \ff+\pf \right)_{,}}^{i} \dd\lambda \right)\Bigg\}. \notag
\end{align}

\subsection{Physical Volume}

The second order perturbation to the physical volume is 
\begin{align}
\label{eq:dvv2}
\dd \dVs &= - \bar{\E}\bar{d}_{A}^{2} \left[ \left(\frac{\daf}{\bar{d}_{A}}\right)^{2} + \left( \frac{\Ef}{\bar{\E}}\right) \left( \frac{\daf}{\bar{d}_{A}}\right) +\frac{\das}{\bar{d}_{A}} +\frac{1}{2}\frac{\Es}{\bar{\E}} \right]\dd \lambda \dd \Omega  \\
&= a^{2}\left(\lambda_{s}\right)\left(\lambda_{s}-\lambda_{o}\right)^{2} \Bigg\{ \Bigg( \ff |_{o} - (v_{1 i}n^{i})_{o}-\ff|_{o}^{s}-\frac{3}{2}\left( \pf-\ff\right)\Big|_{o}^{s} \notag\\
&\qquad +\int_{\lambda_{o}}^{\lambda_{s}}\dd\lambda\int_{\lambda_{s}}^{\lambda_{s}}\dd\tilde{\lambda}\left\{ \left( \ff+\pf\right)'' - \nabla^{2}\left( \ff+\pf \right) \right\} \notag \\
&\qquad -\frac{1}{\lambda_{o}-\lambda_{s}}\left[ 2\int_{\lambda_{o}}^{\lambda_{s}}\dd\lambda\int_{\lambda_{s}}^{\lambda_{s}}\dd\tilde{\lambda}\left( \ff+\pf\right)'-\int_{\lambda_{o}}^{\lambda_{s}}\dd\lambda\int_{\lambda_{s}}^{\lambda_{s}}\dd\tilde{\lambda}\int_{\lambda_{o}}^{\lambda_{s}}\dd\breve{\lambda}\left\{ \left( \ff+\pf \right)''-\nabla^{2}\left( \ff+\pf \right) \right\} \right]\Bigg)^{2}  \notag \\
&\qquad \Bigg(\ff |_{o} - (v_{1 i}n^{i})_{o}-\ff|_{o}^{s}-\frac{3}{2}\left( \pf-\ff\right)\Big|_{o}^{s}+\int_{\lambda_{o}}^{\lambda_{s}}\dd\lambda\int_{\lambda_{s}}^{\lambda_{s}}\dd\tilde{\lambda}\left\{ \left( \ff+\pf\right)'' - \nabla^{2}\left( \ff+\pf \right) \right\} \notag \\
&\qquad -\frac{1}{\lambda_{o}-\lambda{s}}\left[ 2\int_{\lambda_{o}}^{\lambda_{s}}\dd\lambda\int_{\lambda_{s}}^{\lambda_{s}}\dd\tilde{\lambda}\left( \ff+\pf\right)'-\int_{\lambda_{o}}^{\lambda_{s}}\dd\lambda\int_{\lambda_{s}}^{\lambda_{s}}\dd\tilde{\lambda}\int_{\lambda_{o}}^{\lambda_{s}}\dd\breve{\lambda}\left\{ \left( \ff+\pf \right)''-\nabla^{2}\left( \ff+\pf \right) \right\} \right]\Bigg) \times \notag \\
&\qquad \times \Bigg( -2 \ff \Big|_{o}^{s}+\int_{\lambda_{o}}^{\lambda^{s}}\left( \ff' +\pf' \right)\dd\lambda + \ff - v_{1i}n^{i} \Bigg)\notag \\
&\qquad  +\frac{1}{2}\Big[\ff^{2}-5\pf^{2}- 6\ff\pf + \frac{1}{2}\left( \fs-\ps\right) - (v_{1 i}n^{i}) \ff - n^{i}v_{2i} \Big]_{o} \notag\\
&\qquad -\frac{1}{\lambda_{o}-\lambda_{s}}\int_{\lambda_{o}}^{\lambda_{s}}\dd\lambda\int_{\lambda_{s}}^{\lambda_{s}}\dd\tilde{\lambda}\Big\{4\left( -2\ff\Big|^{s}_{s} + \int_{\lambda_{o}}^{\lambda_{s}}\left( \ff'+\pf' \right)\dd \lambda \right)  \Bigg(  \ff |_{o} - (v_{1 i}n^{i})_{o}-\ff|_{o}^{s}\notag \\
&\qquad -\frac{3}{2}\left( \pf-\ff\right)\Big|_{o}^{s}+\int_{\lambda_{o}}^{\lambda_{s}}\dd\lambda\int_{\lambda_{s}}^{\lambda_{s}}\dd\tilde{\lambda}\left\{ \left( \ff+\pf\right)'' - \nabla^{2}\left( \ff+\pf \right) \right\} \notag \\
&\qquad -\frac{1}{\bar{d}_{A}}\left[ 2\int_{\lambda_{o}}^{\lambda_{s}}\dd\lambda\int_{\lambda_{s}}^{\lambda_{s}}\dd\tilde{\lambda}\left( \ff+\pf\right)'-\int_{\lambda_{o}}^{\lambda_{s}}\dd\lambda\int_{\lambda_{s}}^{\lambda_{s}}\dd\tilde{\lambda}\int_{\lambda_{o}}^{\lambda_{s}}\dd\breve{\lambda}\left\{ \left( \ff+\pf \right)''-\nabla^{2}\left( \ff+\pf \right) \right\} \right] \Bigg)'' \notag \\
&\qquad +\Bigg( \ff |_{o} - (v_{1 i}n^{i})_{o}-\ff|_{o}^{s}-\frac{3}{2}\left( \pf-\ff\right)\Big|_{o}^{s}+\int_{\lambda_{o}}^{\lambda_{s}}\dd\lambda\int_{\lambda_{s}}^{\lambda_{s}}\dd\tilde{\lambda}\left\{ \left( \ff+\pf\right)'' - \nabla^{2}\left( \ff+\pf \right) \right\} \notag \\
&\qquad -\frac{1}{\bar{d}_{A}}\left[ 2\int_{\lambda_{o}}^{\lambda_{s}}\dd\lambda\int_{\lambda_{s}}^{\lambda_{s}}\dd\tilde{\lambda}\left( \ff+\pf\right)'-\int_{\lambda_{o}}^{\lambda_{s}}\dd\lambda\int_{\lambda_{s}}^{\lambda_{s}}\dd\tilde{\lambda}\int_{\lambda_{o}}^{\lambda_{s}}\dd\breve{\lambda}\left\{ \left( \ff+\pf \right)''-\nabla^{2}\left( \ff+\pf \right) \right\} \right]\Bigg)' \times \notag \\
&\qquad \times \left( - 2 \frac{\dd \ff}{\dd \lambda} + \ff' + \pf' \right) +\frac{1}{2} \Bigg( \ff |_{o} - (v_{1 i}n^{i})_{o}-\ff|_{o}^{s}-\frac{3}{2}\left( \pf-\ff\right)\Big|_{o}^{s}\notag \\
&\qquad +\int_{\lambda_{o}}^{\lambda_{s}}\dd\lambda\int_{\lambda_{s}}^{\lambda_{s}}\dd\tilde{\lambda}\left\{ \left( \ff+\pf\right)'' - \nabla^{2}\left( \ff+\pf \right) \right\} \notag \\
&\qquad -\frac{1}{\lambda_{o}-\lambda_{s}}\left[ 2\int_{\lambda_{o}}^{\lambda_{s}}\dd\lambda\int_{\lambda_{s}}^{\lambda_{s}}\dd\tilde{\lambda}\left( \ff+\pf\right)'-\int_{\lambda_{o}}^{\lambda_{s}}\dd\lambda\int_{\lambda_{s}}^{\lambda_{s}}\dd\tilde{\lambda}\int_{\lambda_{o}}^{\lambda_{s}}\dd\breve{\lambda}\left\{ \left( \ff+\pf \right)''-\nabla^{2}\left( \ff+\pf \right) \right\} \right]\Bigg) \times\notag \\
&\qquad \Bigg( \left[ \frac{\dd^{2}\pf}{\dd\lambda^{2}}- \frac{\dd^{2}\ff}{\dd\lambda^{2}} \right] +2\left[ \frac{\dd\ff'}{\dd\lambda}+ \frac{\dd\pf'}{\dd\lambda}  \right] + \nabla^{2}\left( \ff+\pf\right) - \left(\ff''+\pf''\right) \Bigg) \Big\}\notag \\
&\qquad +\frac{1}{\lambda_{o}-\lambda_{s}}\int_{\lambda_{o}}^{\lambda_{s}}\dd\lambda\int_{\lambda_{s}}^{\lambda_{s}}\dd\tilde{\lambda}\Bigg\{ \frac{1}{2} \Bigg[ -2 \frac{\dd \fs}{\dd\lambda}+\fs'+\ps' + \frac{1}{2}\left[ \frac{\dd\fs}{\dd\lambda} + \frac{\dd\ps}{\dd\lambda}\right] + 4 \ff \left[3\frac{\dd\ff}{\dd \lambda} - \frac{\dd\pf}{\dd\lambda}\right] \notag \\
&\qquad -4\ff'\left( \ff+\pf\right) + 4\ff \left( 3\ff + \pf \right) + 4 \pf \int_{\lambda_{o}}^{\lambda_{s}}\left( \ff'' + \pf'' \right)\dd \lambda \notag 
\end{align}
\begin{align}
&\qquad -\left( 4\ff + \pf \right)_{,i} \int_{\lambda_{o}}^{\lambda_{s}}{\left( \ff + \pf \right)_{,}}^{i}\dd \lambda + 4 \frac{\dd\pf}{\dd\lambda} \int_{\lambda_{o}}^{\lambda_{s}}\left( \ff' + \pf' \right)\dd \lambda \notag \\
&\qquad -2 \left[ 3\ff + \ff + \ff' + 3 \pf'\right]\int_{\lambda_{o}}^{\lambda_{s}}\left( \ff' + \pf' \right)\dd \lambda\notag \\
&\qquad + 6 \left( \int_{\lambda_{o}}^{\lambda_{s}}\left( \ff + \pf \right)_{,i} \dd \lambda \right) \left( \int_{\lambda_{o}}^{\lambda_{s}} {\left( \ff + \pf \right)_{,}}^{i} \dd \lambda \right) \notag \\
&\qquad +2 \left( \int_{\lambda_{o}}^{\lambda_{s}}\left( \ff' + \pf' \right)\dd \lambda \right) \left( \int_{\lambda_{o}}^{\lambda_{s}}\left( \ff' + \pf' \right)\dd \lambda \right) \Bigg] \notag \\
&\qquad - \left( -2\ff\Big|^{s}_{s} + \int_{\lambda_{o}}^{\lambda_{s}}\left( \ff'+\pf' \right)\dd \lambda \right)\left( -2\ff\Big|^{s}_{s} + \int_{\lambda_{o}}^{\lambda_{s}}\left( \ff'+\pf' \right)\dd \lambda \right)' \notag \\
&\qquad +\left( -2n^{i}\pf\Big|^{s}_{s} + \int_{\lambda_{o}}^{\lambda_{s}}{\left( \ff+\pf \right)_{,}}^{i}\dd \lambda \right)\left( -2\ff\Big|^{s}_{s} + \int_{\lambda_{o}}^{\lambda_{s}}\left( \ff'+\pf' \right)\dd \lambda \right)_{,i}\Bigg\}\notag \\
&\qquad -\frac{1}{2}\int_{\lambda_{o}}^{\lambda_{s}}\dd\lambda\int_{\lambda_{s}}^{\lambda_{s}}\dd\tilde{\lambda}\Bigg\{ \frac{1}{2}\left[ \frac{\dd^{2}\pf}{\dd\lambda^{2}}-\frac{\dd^{2}\ff}{\dd\lambda^{2}} \right] + \left[ \frac{\dd\ff}{\dd\lambda}+\frac{\dd\pf}{\dd\lambda}\right] - \frac{1}{2}\left( \ff'' + \pf'' \right) + \frac{1}{2}\nabla^{2} \left( \fs + \ps \right) \notag \\
&\qquad +\left(\ff'\right)^{2} - \left( \pf '\right)^{2} + 2 \ff \left[ \ff''-\nabla^{2}\ff - \frac{\dd^{2}\pf}{\dd\lambda^{2}} - 3 \frac{\dd\ff'}{\dd\lambda} \right] \notag \\
&\qquad - 2 \pf \left[ \pf''-\nabla^{2}\pf - \frac{\dd^{2}\ff}{\dd\lambda^{2}} - 3 \frac{\dd\pf'}{\dd\lambda} \right] -\ff_{,i}{\ff_{,}}^{i}+\pf_{,i}{\pf_{,}}^{i}\notag \\
&\qquad +4\pf\left[ \frac{\dd^{2}\pf}{\dd\lambda^{2}} - \frac{\dd^{2}\ff}{\dd\lambda^{2}} +\frac{\dd \ff'}{\dd\lambda} - \frac{\dd \pf'}{\dd\lambda} + \frac{\dd \ff}{\dd\lambda}+ \frac{\dd \pf}{\dd\lambda} -\ff'-\pf' \right] \notag\\
&\qquad -2 \left[ \frac{\dd \pf}{\dd\lambda} - \frac{\dd \ff}{\dd\lambda} +\ff + \pf \right]_{,i}\int_{\lambda_{o}}^{\lambda_{s}}{\left( \ff + \pf\right)_{,}}^{i} + 4\ff \frac{\dd\pf}{\dd\lambda} + 4\pf \frac{\dd\pf}{\dd\lambda} \notag \\
&\qquad +2 \left( \ff+\pf\right)\nabla^{2}\left( \ff+\pf \right) + 4\frac{\dd\pf}{\dd\lambda}\left[ -2\ff\Big|^{s}_{o} + \int_{\lambda_{o}}^{\lambda_{s}}\left( \ff' + \pf'\right)\dd\lambda\right] \notag \\
&\qquad + 2 \nabla^{2}\left( \ff+\pf\right)\left[ -2\ff\Big|^{s}_{o} + \int_{\lambda_{o}}^{\lambda_{s}}\left( \ff' + \pf'\right)\dd\lambda\right]\notag \\
&\qquad -\frac{3}{2} \Bigg[ \frac{\dd}{\dd \lambda}\left( \ff + \pf \right) -2 \left( \ff' + \pf' \right) + \int_{\lambda_{o}}^{\lambda_{s}}\left( \ff''+\pf'' \right)\dd\lambda \Bigg]^{2} \notag \\
& \qquad + 8 \left( \ff' + \pf' \right)^{2} + 2 \left[ \int_{\lambda_{o}}^{\lambda_{s}}\left( \ff'' + \pf'' \right)\dd\lambda\right]^{2} + 2 \left[ \frac{\dd \ff}{\dd\lambda} + \frac{\dd \pf}{\dd \lambda} \right]^{2}\notag \\
&\qquad - \left( \ff+\pf \right) \Bigg[ 8\frac{\dd}{\dd\lambda}\left( \ff'' + \pf'' \right) + 4\left( \ff'' + \pf'' \right) + \nabla^{2}\left( \ff + \pf \right)  \Bigg] \notag \\
&\qquad - \left( \ff' + \pf' \right) \int_{\lambda_{o}}^{\lambda_{s}}\Big[ 8\left( \ff'' + \pf'' \right)-2\nabla^{2}\left( \ff + \pf \right) \Big]\dd\lambda \notag \\
&\qquad +\left(\int_{\lambda_{o}}^{\lambda_{s}}\left( \ff+\pf \right)_{,ij}\dd\lambda \right) \left(\int_{\lambda_{o}}^{\lambda_{s}}{\left( \ff+\pf \right)_{,}}^{ij}\dd\lambda\right) \notag \\
&\qquad +\left(\int_{\lambda_{o}}^{\lambda_{s}}\nabla^{2}\left( \ff+\pf \right)\dd\lambda \right) \left(\int_{\lambda_{o}}^{\lambda_{s}}\left( \ff'' + \pf'' \right)\dd\lambda\right) \notag \\
&\qquad -\left(\int_{\lambda_{o}}^{\lambda_{s}}\nabla^{2}\left( \ff+\pf \right)\dd\lambda \right) \left(\int_{\lambda_{o}}^{\lambda_{s}}\nabla^{2}\left( \ff + \pf \right)\dd\lambda\right) \notag \\
&\qquad - 2\left( \ff+\pf \right)_{,i}{\left( \ff+\pf \right)_{,}}^{i}  + 4 {\left( \ff+\pf \right)_{,}}^{i} \Bigg[ \int_{\lambda_{o}}^{\lambda_{s}}\left( \ff+\pf \right)_{,i} \dd\lambda\Bigg]\notag \\
&\qquad - 2 \left( \int_{\lambda_{o}}^{\lambda_{s}}\left( \ff+\pf \right)_{,i} \dd\lambda \right)\left( \int_{\lambda_{o}}^{\lambda_{s}}{\left( \ff+\pf \right)_{,}}^{i} \dd\lambda \right)\Bigg\}\notag 
\end{align}
\begin{align}
&\qquad+\frac{2}{\lambda_{o}-\lambda_{s}} \int_{\lambda_{o}}^{\lambda_{s}}\dd\lambda\int_{\lambda_{s}}^{\lambda_{s}}\dd\tilde{\lambda}\int_{\lambda_{o}}^{\lambda_{s}}\dd\breve{\lambda}\Bigg\{ \frac{1}{2}\left[ \frac{\dd^{2}\pf}{\dd\lambda^{2}}-\frac{\dd^{2}\ff}{\dd\lambda^{2}} \right] + \left[ \frac{\dd\ff}{\dd\lambda}+\frac{\dd\pf}{\dd\lambda}\right] - \frac{1}{2}\left( \ff'' + \pf'' \right) + \frac{1}{2}\nabla^{2} \left( \fs + \ps \right) \notag \\
&\qquad +\left(\ff'\right)^{2} - \left( \pf '\right)^{2} + 2 \ff \left[ \ff''-\nabla^{2}\ff - \frac{\dd^{2}\pf}{\dd\lambda^{2}} - 3 \frac{\dd\ff'}{\dd\lambda} \right] \notag \\
&\qquad - 2 \pf \left[ \pf''-\nabla^{2}\pf - \frac{\dd^{2}\ff}{\dd\lambda^{2}} - 3 \frac{\dd\pf'}{\dd\lambda} \right] -\ff_{,i}{\ff_{,}}^{i}+\pf_{,i}{\pf_{,}}^{i}\notag \\
&\qquad +4\pf\left[ \frac{\dd^{2}\pf}{\dd\lambda^{2}} - \frac{\dd^{2}\ff}{\dd\lambda^{2}} +\frac{\dd \ff'}{\dd\lambda} - \frac{\dd \pf'}{\dd\lambda} + \frac{\dd \ff}{\dd\lambda}+ \frac{\dd \pf}{\dd\lambda} -\ff'-\pf' \right] \notag\\
&\qquad -2 \left[ \frac{\dd \pf}{\dd\lambda} - \frac{\dd \ff}{\dd\lambda} +\ff + \pf \right]_{,i}\int_{\lambda_{o}}^{\lambda_{s}}{\left( \ff + \pf\right)_{,}}^{i} + 4\ff \frac{\dd\pf}{\dd\lambda} + 4\pf \frac{\dd\pf}{\dd\lambda} \notag \\
&\qquad +2 \left( \ff+\pf\right)\nabla^{2}\left( \ff+\pf \right) + 4\frac{\dd\pf}{\dd\lambda}\left[ -2\ff\Big|^{s}_{o} + \int_{\lambda_{o}}^{\lambda_{s}}\left( \ff' + \pf'\right)\dd\lambda\right] \notag \\
&\qquad + 2 \nabla^{2}\left( \ff+\pf\right)\left[ -2\ff\Big|^{s}_{o} + \int_{\lambda_{o}}^{\lambda_{s}}\left( \ff' + \pf'\right)\dd\lambda\right] \notag \\
&\qquad -\frac{3}{2} \Bigg[ \frac{\dd}{\dd \lambda}\left( \ff + \pf \right) -2 \left( \ff' + \pf' \right) + \int_{\lambda_{o}}^{\lambda_{s}}\left( \ff''+\pf'' \right)\dd\lambda \Bigg]^{2} \notag \\
& \qquad + 8 \left( \ff' + \pf' \right)^{2} + 2 \left[ \int_{\lambda_{o}}^{\lambda_{s}}\left( \ff'' + \pf'' \right)\dd\lambda\right]^{2} + 2 \left[ \frac{\dd \ff}{\dd\lambda} + \frac{\dd \pf}{\dd \lambda} \right]^{2}\notag \\
&\qquad - \left( \ff+\pf \right) \Bigg[ 8\frac{\dd}{\dd\lambda}\left( \ff'' + \pf'' \right) + 4\left( \ff'' + \pf'' \right) + \nabla^{2}\left( \ff + \pf \right)  \Bigg] \notag \\
&\qquad - \left( \ff' + \pf' \right) \int_{\lambda_{o}}^{\lambda_{s}}\Big[ 8\left( \ff'' + \pf'' \right)-2\nabla^{2}\left( \ff + \pf \right) \Big]\dd\lambda \notag \\
&\qquad +\left(\int_{\lambda_{o}}^{\lambda_{s}}\left( \ff+\pf \right)_{,ij}\dd\lambda \right) \left(\int_{\lambda_{o}}^{\lambda_{s}}{\left( \ff+\pf \right)_{,}}^{ij}\dd\lambda\right) \notag \\
&\qquad +\left(\int_{\lambda_{o}}^{\lambda_{s}}\nabla^{2}\left( \ff+\pf \right)\dd\lambda \right) \left(\int_{\lambda_{o}}^{\lambda_{s}}\left( \ff'' + \pf'' \right)\dd\lambda\right) \notag \\
&\qquad -\left(\int_{\lambda_{o}}^{\lambda_{s}}\nabla^{2}\left( \ff+\pf \right)\dd\lambda \right) \left(\int_{\lambda_{o}}^{\lambda_{s}}\nabla^{2}\left( \ff + \pf \right)\dd\lambda\right) \notag \\
&\qquad - 2\left( \ff+\pf \right)_{,i}{\left( \ff+\pf \right)_{,}}^{i}  + 4 {\left( \ff+\pf \right)_{,}}^{i} \Bigg[ \int_{\lambda_{o}}^{\lambda_{s}}\left( \ff+\pf \right)_{,i} \dd\lambda\Bigg]\notag \\
&\qquad - 2 \left( \int_{\lambda_{o}}^{\lambda_{s}}\left( \ff+\pf \right)_{,i} \dd\lambda \right)\left( \int_{\lambda_{o}}^{\lambda_{s}}{\left( \ff+\pf \right)_{,}}^{i} \dd\lambda \right)\Bigg\} \notag \\
&\qquad +\frac{1}{4}\Bigg[ 2 \Big\{  \left(-2 \ff \Big|_{o}^{s}+\int_{\lambda_{o}}^{\lambda^{s}}\left( \ff' +\pf' \right)\dd\lambda\right) + \ff - v_{1i}n^{i}  \Big\} \ff \notag \\
&\qquad - \ff^{2}-5\pf^{2}- 6\ff\pf + \frac{1}{2}\left( \fs-\ps\right) - n^{i}v_{2i} \notag \\
&\qquad - 4\left( \ff + \pf\right) \ff\Big|^{s}_{o}+ 2\left( \ff + \pf\right)\int_{\lambda_{o}}^{\lambda_{s}}\left( \ff' + \pf' \right)\dd\lambda  \notag \\
&\qquad  -4n^{i}v_{1 i} \pf\Big|^{s}_{o} +2 v_{1 i}  \int_{\lambda_{o}}^{\lambda_{s}}{\left( \ff + \pf \right)_{,}}^{i} \dd\lambda \notag \\
&\qquad +\int_{\lambda_{o}}^{\lambda_{s}}\Bigg\{-2 \frac{\dd \fs}{\dd\lambda}+\fs'+\ps' + \frac{1}{2}\left[ \frac{\dd\fs}{\dd\lambda} + \frac{\dd\ps}{\dd\lambda}\right] + 4 \ff \left[3\frac{\dd\ff}{\dd \lambda} - \frac{\dd\pf}{\dd\lambda}\right] \notag 
\end{align}
\begin{align}
&\qquad -4\ff'\left( \ff+\pf\right) + 4\ff \left( 3\ff + \pf \right) + 4 \pf \int_{\lambda_{o}}^{\lambda_{s}}\left( \ff'' + \pf'' \right)\dd \lambda \notag \\
&\qquad -\left( 4\ff + \pf \right)_{,i} \int_{\lambda_{o}}^{\lambda_{s}}{\left( \ff + \pf \right)_{,}}^{i}\dd \lambda + 4 \frac{\dd\pf}{\dd\lambda} \int_{\lambda_{o}}^{\lambda_{s}}\left( \ff' + \pf' \right)\dd \lambda \notag \\
&\qquad -2 \left[ 3\ff + \ff + \ff' + 3 \pf'\right]\int_{\lambda_{o}}^{\lambda_{s}}\left( \ff' + \pf' \right)\dd \lambda\notag \\
&\qquad + 6 \left( \int_{\lambda_{o}}^{\lambda_{s}}\left( \ff + \pf \right)_{,i} \dd \lambda \right) \left( \int_{\lambda_{o}}^{\lambda_{s}} {\left( \ff + \pf \right)_{,}}^{i} \dd \lambda \right) \notag \\
&\qquad +2 \left( \int_{\lambda_{o}}^{\lambda_{s}}\left( \ff' + \pf' \right)\dd \lambda \right) \left( \int_{\lambda_{o}}^{\lambda_{s}}\left( \ff' + \pf' \right)\dd \lambda \right)\Bigg\} \dd \tilde{\lambda} \Bigg]  \Bigg\} \dd \lambda \dd \Omega. \notag
\end{align}

\subsection{Redshift Density}

Using \eqs{eq:dz1}, \eqref{eq:dz2} and \eqref{eq:der2-rho} in \eq{eq:redshift-density}  we find that the redshift density perturbation at second order is given by
\begin{align}
\label{eq:ddz2}
\delta^{(2)}_{z}(n^{i},z) &= \frac{1}{2} \frac{\drs(n^{i},z)}{\bar{\rho(z)}} + \frac{3}{2(1+\bar{z})} \dzs(n^{i},z) + \frac{3}{(1+z)^{2}}\left[ \dzf(n^{i},z) \right]^{2}  \\
&=  \frac{1}{2} \frac{\drs(n^{i},z)}{\bar{\rho(z)}} + \frac{3}{2(1+\bar{z})} \Bigg( \left[ \ff^{2} + \left( v_{1i}n^{i}\right)^{2} -2 \ff\left( v_{1i}n^{i}\right) \right]_{o}\notag \\
&\qquad -\ff|_{o} \left[ -2\ff + \int_{\lambda_{o}}^{\lambda_{s}}\left( \ff' + \pf' \right)\dd\lambda \right]_{s} \notag\\
&\qquad + \left[ -2\ff + \int_{\lambda_{o}}^{\lambda_{s}}\left( \ff' + \pf' \right)\dd\lambda \right]_{s} \left( v_{1i}n^{i}\right)_{o} -\ff|_{s} \ff|_{o} \notag \\
&\qquad + \ff|_{s}\left( v_{1i}n^{i}\right)_{o}+\left( v_{1i}n^{i}\right)_{s}\ff|_{o}-\left( v_{1i}n^{i}\right)_{s}\left( v_{1i}n^{i}\right)_{o} \notag \\
&\qquad + \frac{1}{2}\Bigg[ 2 \Big\{  \left(-2 \ff \Big|_{o}^{s}+\int_{\lambda_{o}}^{\lambda^{s}}\left( \ff' +\pf' \right)\dd\lambda\right) + \ff - v_{1i}n^{i}  \Big\} \ff \notag \\
&\qquad - \ff^{2}-5\pf^{2}- 6\ff\pf + \frac{1}{2}\left( \fs-\ps\right) - n^{i}v_{2i} \notag \\
&\qquad - 4\left( \ff + \pf\right) \ff\Big|^{s}_{o}+ 2\left( \ff + \pf\right)\int_{\lambda_{o}}^{\lambda_{s}}\left( \ff' + \pf' \right)\dd\lambda  \notag \\
&\qquad  -4n^{i}v_{1 i} \pf\Big|^{s}_{o} +2 v_{1 i}  \int_{\lambda_{o}}^{\lambda_{s}}{\left( \ff + \pf \right)_{,}}^{i} \dd\lambda \notag \\
&\qquad +\int_{\lambda_{o}}^{\lambda_{s}}\Bigg\{-2 \frac{\dd \fs}{\dd\lambda}+\fs'+\ps' + \frac{1}{2}\left[ \frac{\dd\fs}{\dd\lambda} + \frac{\dd\ps}{\dd\lambda}\right] + 4 \ff \left[3\frac{\dd\ff}{\dd \lambda} - \frac{\dd\pf}{\dd\lambda}\right] \notag \\
&\qquad -4\ff'\left( \ff+\pf\right) + 4\ff \left( 3\ff + \pf \right) + 4 \pf \int_{\lambda_{o}}^{\lambda_{s}}\left( \ff'' + \pf'' \right)\dd \lambda \notag \\
&\qquad -\left( 4\ff + \pf \right)_{,i} \int_{\lambda_{o}}^{\lambda_{s}}{\left( \ff + \pf \right)_{,}}^{i}\dd \lambda + 4 \frac{\dd\pf}{\dd\lambda} \int_{\lambda_{o}}^{\lambda_{s}}\left( \ff' + \pf' \right)\dd \lambda \notag \\
&\qquad -2 \left[ 3\ff + \ff + \ff' + 3 \pf'\right]\int_{\lambda_{o}}^{\lambda_{s}}\left( \ff' + \pf' \right)\dd \lambda\notag \\
&\qquad + 6 \left( \int_{\lambda_{o}}^{\lambda_{s}}\left( \ff + \pf \right)_{,i} \dd \lambda \right) \left( \int_{\lambda_{o}}^{\lambda_{s}} {\left( \ff + \pf \right)_{,}}^{i} \dd \lambda \right) \notag \\
&\qquad +2 \left( \int_{\lambda_{o}}^{\lambda_{s}}\left( \ff' + \pf' \right)\dd \lambda \right) \left( \int_{\lambda_{o}}^{\lambda_{s}}\left( \ff' + \pf' \right)\dd \lambda \right)\Bigg\} \dd \tilde{\lambda} \Bigg]_{s} \notag\\
&\qquad - \frac{1}{2}\Bigg[ \ff^{2}-5\pf^{2}- 6\ff\pf + \frac{1}{2}\left( \fs-\ps\right) - (v_{1 i}n^{i}) \ff - n^{i}v_{2i}  \Bigg]_{o} \Bigg) \notag \\
& \quad + \frac{3}{(1+z)^{2}}\left\{ \left( v_{1i}n^{i} - \ff\right)\big|^{s}_{o} + \int_{\lambda_{o}}^{\lambda_{s}}\dd\lambda  \left[\ff' +\pf'\right] \right\}^{2} \notag \\
&\qquad + \frac{1}{{\cal{H}}(1+\bar{z})}\left( \frac{\dd}{\dd \eta}\left( v_{1i}n^{i}\right) - \left[\ff' +\pf'\right] \right) \left( \left( v_{1i}n^{i} - \ff\right)\big|^{s}_{o} + \int_{\lambda_{o}}^{\lambda_{s}}\dd\lambda  \left[\ff' +\pf'\right] \right) \notag,
\end{align}

{
\bibliographystyle{ieeetr}
\bibliography{mybib2}
}

\end{document}